*Phylogenetics*

# Building alternative consensus trees and supertrees using *k*-means and Robinson and Foulds distance

Nadia Tahiri[1,2], Bernard Fichet[3] and Vladimir Makarenkov[1*]

[1]Département d'informatique, Université du Québec à Montréal, Montreal, QC, Canada

[2]Département d'informatique, Université de Sherbrooke, Sherbrooke, QC, Canada

[3]Aix-Marseille Université, Faculté de Médecine, 27 Bd. Jean Moulin, F-13385 Marseille, France

*To whom correspondence should be addressed.

## Abstract

**Motivation:** Each gene has its own evolutionary history which can substantially differ from evolutionary histories of other genes. For example, some individual genes or operons can be affected by specific horizontal gene transfer or recombination events. Thus, the evolutionary history of each gene should be represented by its own phylogenetic tree which may display different evolutionary patterns from the species tree that accounts for the main patterns of vertical descent. However, the output of traditional consensus tree or supertree inference methods is a unique consensus tree or supertree.

**Results:** We present a new efficient method for inferring multiple alternative consensus trees and supertrees to best represent the most important evolutionary patterns of a given set of gene phylogenies. We show how an adapted version of the popular *k*-means clustering algorithm, based on some remarkable properties of the Robinson and Foulds distance, can be used to partition a given set of trees into one (for homogeneous data) or multiple (for heterogeneous data) cluster(s) of trees. Moreover, we adapt the popular Caliński-Harabasz, Silhouette, Ball and Hall, and Gap cluster validity indices to tree clustering with *k*-means. Special attention is given to the relevant but very challenging problem of inferring alternative supertrees. The use of the Euclidean property of the objective function of the method makes it faster than the existing tree clustering techniques, and thus better suited for analyzing large evolutionary datasets.

**Availability and implementation:** Our *KMeansSuperTreeClustering* program along with its C++ source code is available at: https://github.com/TahiriNadia/KMeansSuperTreeClustering.

**Contact:** makarenkov.vladimir@uqam.ca

**Supplementary information:** Supplementary data are available at *Bioinformatics* online.

## 1  Introduction

Most of the conventional consensus and supertree inference methods generate one candidate tree for a given set of input gene phylogenies. However, the topologies of gene phylogenies can be substantially different from each other due to possible horizontal gene transfer, hybridization or intergenic/intergenic recombination events by which the evolution of the related genes could be affected (Bapteste *et al.* 2004). Each gene phylogeny depicts a unique evolutionary history which does not always coincide with the main patterns of vertical descent depicted by the species tree. In order to infer a reliable species phylogeny, the related gene trees should be merged, while minimizing topological conflicts present in them (Maddison *et al.* 2007). In this context, two scenarios can be envisaged: First, trees to be merged are constructed for the same set of taxa (i.e. consensus tree issue), and second, trees to be merged are constructed for different, but mutually overlapping, sets of taxa (i.e. supertree issue).

A large variety of methods have been proposed to address the problem of reconciliation of multiple trees defined on the same set of leaves in order to infer a consensus tree. The best known examples of consensus trees are the strict consensus tree, the majority-rule consensus tree, and the extended majority-rule consensus tree (Bryant 2003). However, in practical evolutionary studies, we rarely deal with phylogenies defined on the same set of taxa, and thus the consensus tree inference problem is transformed into the supertree inference problem (Bininda-Emonds 2004). Several approaches have been proposed to synthesize collections of small phylogenetic trees with partial taxon overlap into comprehensive supertrees including all taxa found in the input trees (Wilkinson *et al.* 2007). For example, the Matrix Representation with Parsimony (MRP) methods (Baum 1992; Ragan 1992) proceed by a matrix-like aggregation of separately inferred partial trees. The trees derived from these independent analyses are then combined to produce a single MRP matrix used to reconstruct the supertree of all sources of taxa (de Queiroz and Gatesy 2007). In the parsimony supermatrix methods all systematic characters are integrated into a single phylogenetic matrix which is used to analyze all characters simultaneously in order to build a supermatrix tree. The strict supertree includes the bipartitions (or splits) that agree with all bipartitions present in the input phylogenies, while the rest of the tree consists of unresolved subtrees. The concept of strict and loose supertrees has been well described by McMorris and Wilkinson (2011), who showed that these types of supertrees are natural generalizations of the corresponding consensus trees. Importantly, most supertree methods are heuristics for NP-hard optimization problems (Warnow 2018).

The implementation of the famous "Tree of Life" (ToL) project intended to infer the largest possible species phylogeny became feasible due to collaborative efforts of biologists and nature enthusiasts from around the world (Maddison *et al.* 2007). The approach adopted by the project organizers consists in gradual partitioning the complex tree reconstruction problem into several subproblems, followed by merging the obtained sub-trees. Indeed, such an approach produces thousands of small trees which should be assembled to create ToL. Precisely, the problem can be viewed as twofold: First, we have to infer small sub-trees of ToL (i.e. often gene trees representing different evolutionary histories), the most commonly defined on different, but mutually overlapping, sets of taxa, and second, we have to merge these small sub-trees into one or several supertrees using a supertree reconstruction method that allows combining trees inferred for different sets of taxa. In this context, the application of a method that infers multiple supertrees (i.e. a supertree clustering method) would help discover and regroup plausible alternative evolutionary scenarios for several sub-trees of ToL.

Unfortunately, most of the traditional consensus tree and supertree inference methods return as output a single consensus tree or supertree. Thus, in many instances, these methods are not informative enough, as they do not preserve alternative evolutionary scenarios characterizing sub-groups of genes that have undergone similar reticulate evolutionary events (e.g. genes whose evolution has been affected by similar horizontal gene transfers; Boc *et al.* 2010; Sevillya *et al.* 2020). The Phylogenetics Islands method of Maddison (1991) was the first to use multiple consensus trees. Maddison observed that the consensus trees of the islands can differ substantially from each other, and that they are usually much better resolved than the consensus tree of the whole set of taxa. The most intuitive approach for discovering and regrouping genes sharing similar evolutionary histories is clustering their gene phylogenies. In this context, Stockham *et al.* (2002) proposed a tree clustering algorithm based on $k$-means (MacQueen 1967) and the quadratic version of the Robinson and Foulds ($RF$) topological distance (Robinson and Foulds 1981). The $RF$ distance between two phylogenetic trees equals the minimum number of elementary operations, consisting of merging or splitting nodes, which are necessary to transform one tree into the other. It is also the number of bipartitions, or Buneman's splits (Buneman 1971), which belong to exactly one of the two trees. The clustering algorithm proposed by Stockham *et al.* aims at inferring a set of strict consensus trees that minimizes the information loss. It proceeds by determining the consensus trees for each set of clusters in all intermediate partitioning solutions tested by $k$-means. Bonnard *et al.* (2006) introduced a method, called MultiPolar Consensus (MPC), which produces a minimum number of consensus trees displaying all splits of a given set of phylogenies whose support is above a predefined threshold. Guénoche (2013) proposed the Multiple Consensus Trees (MCT) method for partitioning a group of phylogenetic trees into one or several clusters. The method of Guénoche computes a generalized partition score to determine the most appropriate number of clusters for a given set of gene trees. Finally, Tahiri *et al.* (2018) presented a fast tree clustering method based on $k$-medoids. Recently, Silva and Wilkinson (2021) have introduced a revised definition of tree islands based on any tree-to-tree metric that usefully extends this notion to any set or multiset of trees and have provided a nice discussion of biological applications of their method. To the best of our knowledge, all the methods proposed for building multiple alternative phylogenies assume that the input trees have identical sets of taxa, thus working in the consensus tree context. Therefore, the relevant and challenging problem of inferring multiple supertrees still needs to be addressed appropriately. It is worth noting that Bansal *et al.* (2010) described a method for building $RF$-based supertrees aiming at minimizing the total $RF$ distance between the supertree and the set of input trees. However, Bansal *et al.* did not consider the possibility of inferring multiple supertrees.

In this paper, we describe a new method that can be used to infer all alternative consensus trees and supertrees. We present some interesting properties of the Robinson and Foulds topological distance and show that it should not be used in its quadratic form in tree clustering. Our main contributions are summarized below:

1. We present a new efficient method for inferring multiple alternative consensus trees which is based on the Euclidean properties of the square root of the Robinson and Foulds distance, and validate it through a comprehensive simulation study;

2. We show how the proposed method can be extended to infer multiple alternative supertrees (this problem has not been addressed yet in the literature);

3. We demonstrate the practical utility of our method and software by applying them to the problem of clustering SARS-Cov-2 gene trees and by comparing our results to those provided by existing tree clustering approaches;

4. We show how to adapt the popular Ball and Hall (Ball-Hall 1967), Caliński-Harabasz (Caliński and Harabasz 1974), Gap (Tibshirani *et al.* 2001) and Silhouette (Rousseeuw 1987) cluster validity indices to tree clustering with $k$-means;

5. We discuss further extensions of the proposed tree and supertree clustering approach.

## 2    Methods

A phylogenetic tree is an unrooted leaf-labeled tree in which each internal node, representing an ancestor of some contemporary species (i.e. taxa), has at least two children and all leaves, representing contemporary species, have different labels. Our method takes as input a set $\Pi$ of $N$ phylogenetic trees defined on the same (or different, but mutually overlapping) set(s) of leaves and returns as output one or several disjoint clusters of trees from which the related consensus trees (or supertrees) can then be inferred. Our approach relies of the use of a fast version of the popular $k$-means algorithm adapted for tree clustering. We define and compare different variants of the $k$-means objective function suitable for clustering trees.

$K$-means (MacQueen 1967) is an unsupervised data partitioning algorithm which iteratively regroups $N$ given objects (i.e. phylogenetic trees in our case) into $K$ disjoint clusters. The content of each cluster is chosen to minimize the sum of intracluster distances. The most commonly used distances in the framework of $k$-means are the Euclidean distance, the Manhattan distance, and the Minkowski distance. The problem of finding an optimal partitioning according to the $k$-means least-squares criterion is known to be NP-hard (Mahajan *et al.* 2009). This fact has motivated the development of polynomial-time heuristics for finding an approximate clustering solution, most of them having the time complexity of $O(KNIM)$, where $I$ is the number of iterations in $k$-means and $M$ is the number of variables characterizing each of the $N$ objects.

The Robinson and Foulds distance, also known as the symmetric-difference distance, between two trees is a widely-used metric for comparison of phylogenetic trees defined on the same set of taxa (Robinson and Foulds 1981). The $RF$ distance is a topological distance. It does not take into account the length of the tree edges. As shown by Barthélemy and McMorris (1986), the majority-rule consensus tree of a set of trees is a median tree of this set in the sense of the $RF$ distance. Moreover, even though the $RF$ distance itself is not a Euclidean distance, its square root is Euclidean (see Appendix S1 in Supplementary Material). This fact justifies the use of this distance in tree clustering.

One of the main advantages of the method described in this paper is that it does not recompute the consensus trees or supertrees for all intermediate clusters of trees, but estimates the quality of each intermediate tree clustering using formulas based on the properties of the $RF$ distance and majority-rule consensus trees. This allows for a much faster clustering of a given set of input phylogenies without compromising on the quality of the resulting consensus trees or supertrees.

### 2.1    Tree clustering problem

The traditional $k$-means algorithm (MacQueen 1967) partitions a given dataset into $K$ ($K > 1$) disjoint clusters according to its objective function based on a specific distance (e.g. Euclidean or Minkowski distance) and the selected cluster validity index (e.g. Caliński-Harabasz, Silhouette or Dann index). Most of the traditional cluster validity indices take into consideration both intragroup and intergroup cluster evaluations. However, we cannot use the standard objective functions or cluster validity indices when clustering trees. Here, we discuss the main modifications that should be introduced into the conventional $k$-means algorithm in order to adapt it to tree clustering.

In case of tree clustering, the objective function of the method can be defined as follows:

$$OF = \sum_{k=1}^{K} \sum_{i=1}^{N_k} RF(C_k, T_{ki}), \quad (1)$$

where $K$ is the number of tree clusters, $N_k$ is the number of trees in cluster $k$, $RF(C_k, T_{ki})$ is the RF distance between tree $i$ of cluster $k$, denoted $T_{ki}$, and the majority-rule consensus tree (any other type of consensus tree could be considered here as well) of cluster $k$, denoted $C_k$. If the majority-rule consensus tree is used in the objective function of the tree clustering algorithm, the problem is sometimes called the $k$-median tree clustering problem. Since we do not compute any median tree during the clustering process, we present it as the $k$-means tree clustering problem. This notation was also used in the pioneering work of Stockham *et al.* (2002), who considered the consensus tree of each cluster as its mean.

Still, the computation of the majority-rule, or of the extended majority-rule, consensus tree is time-consuming. The running time of the straightforward algorithm computing any of these consensus trees is $O(n^2 + nN^2)$ (Wareham 1985), where $n$ is the number of leaves in each tree and $N$ is the number of trees. If the trees are defined by their sets of weighted branches, an optimal algorithm proposed by Jansson *et al.* (2013), and running in $O(nN)$ time, can be used for computing the majority-rule consensus tree. The time complexity of the version of the tree partitioning algorithm using consensus trees as cluster centers (Equation 1) which was tested in our simulations is $O(rmN^2KI)$ (*i.e.* it is a variant of the algorithm by Stockham *et al.*), where $r$ is the number of different random starting partitions used in $k$-means (typically, hundreds of different starting partitions should be considered in order to achieve a good clustering performance with $k$-means, see Steinley and Brusco (2007)), and $I$ is the number of iterations in the internal loop of $k$-means. In our simulations, we compared different versions of this algorithm which are as follows:

a) The version that uses strict consensus trees (as in Stockham *et al.*) and the version that uses majority-rule consensus trees - the latter provided better results;

b) The version that uses different random starts (*i.e.* $r=100$) and the version that uses only one random starting partition (*i.e.* $r=1$) - the former provided much better results;

c) The version that uses the first-improvement strategy, i.e. any improving relocation is applied without updating the cluster centers (i.e. without recalculating consensus trees of the clusters), and the version that uses the best-improvement strategy when all possible tree relocations are iteratively tested and only the best one is applied - the latter provided better results.

Thus, the time complexity of the consensus tree-based $k$-means that favors speed is $O(nNKI)$, while the time complexity of the version that favors solution quality, at the expense of extra computational time, is $O(rmN^2KI)$ (here, we consider that $O(nN^2K)$ time is needed to relocate $N$ trees).

It is worth mentioning that Tahiri *et al.* (2018) have recently proposed a $k$-medoids-based tree clustering algorithm having the running time of $O(nN^2 + rK(N-K)^2I)$, where $O(nN^2)$ is the time needed to precalculate the matrix of pairwise RF distances of size ($N \times N$) between all trees in $\Pi$.

In order to speed up the tree clustering process, we propose to use the following objective function $OF_{EA}$:

$$OF_{EA} = SS_W = \sum_{k=1}^{K} \sum_{i=1}^{N_k} \left\| T_{ki}^v - m_k^v \right\|^2 = \sum_{k=1}^{K} \frac{1}{N_k} \sum_{i=1}^{N_k-1} \sum_{j=i+1}^{N_k} RF(T_{ki}, T_{kj}), \quad (2)$$

where $SS_W$ is the index of intragroup evaluation, $T_{ki}^v$ and $m_k^v$ are, respectively, the tree $i$ of cluster $k$ and the overall mean (i.e. centroid) of cluster $k$ in the *tree vector space*. In fact, trees can be represented as vectors of their bipartitions or splits. There exist $2^{n-1}-1$ possible splits for $n$-leaf trees, and at most $2n-3$ splits can occur in a tree. In a tree vector representation, all coordinates that do not correspond to a split of the trees are set to 0; those that correspond to a split are set to 1 (for more details, see St. John 2017). In Equation 2, the RF distance, or more precisely its square root, which has the Euclidean property, replaces the traditional Euclidean distance used in $k$-means (see also Equation 15 in Appendix S2 in Supplementary Material that provides two equivalent expressions for $SS_W$ for the traditional Euclidean distance). Importantly, the centroid vector $m_k^v$ of a cluster of trees $k$ is not necessarily a consensus tree of the cluster. Moreover, $m_k^v$ may not correspond to a phylogenetic tree at all. Fortunately, we do not need to calculate the $L^2$ norm (the middle term in Equation 2) or work with $2^{n-1}-1$-dimentional vectors in the clustering process. The main advantage of using the precalculated RF distances (the right term in Equation 2) is that we should not calculate a consensus tree, or the cluster centroid, for any intermediate cluster of trees considered by $k$-means, and that an object relocation operation consisting in finding the best cluster for a given tree $T$ belonging to cluster $C$ (i.e. when we try to relocate $T$ into each of the $K$-1 clusters that are different from $C$) can be performed in $O(K)$ time. Indeed, all pairwise RF distances between trees in a given set of input trees $\Pi$ can be precalculated, and the sums of the RF distances between a given tree $T$ and all elements of any tree cluster can first be precalculated and then updated at each step of $k$-means. Hence, using Equation 2, an object relocation operation conducted in turn for all $N$ objects (i.e. trees) takes $O(NK)$ time. The RF distance precalculation step can be completed in $O(nN^2)$ (Makarenkov and Leclerc 2000; Sul and Williams 2008) and should be carried out only once for all random starting partitions used in $k$-means. This leads to the total running time of $O(nN^2 + rNKI)$ for our extended majority-rule $k$-means based on Equation 2. Our clustering approach is especially efficient when the number of leaves $n$ is larger than the number of trees $N$. It is worth noting that the objective function $OF_{EA}$ defined in Equation 2 *can also be viewed as a Euclidean approximation of the objective function OF* defined in Equation 1.

Importantly, Equation 2 can also be used to define a version of the Caliński-Harabasz cluster validity index (*CH*) (Caliński and Harabasz 1974) adapted for clustering trees. The *CH* index, which is sometimes called the variance ratio criterion, is defined as follows:

$$CH = \frac{SS_B}{SS_W} \times \frac{N-K}{K-1}, \quad (3)$$

where $SS_B$ is the index of intergroup evaluation, $SS_W$ is the index of intragroup evaluation, $K$ is the number of clusters, and $N$ is the number of objects (i.e. trees in our case). The optimal number of clusters corresponds to the largest value of *CH*.

In the case of tree clustering, the global variance between groups, $SS_B$, can be calculated as follows:

$$SS_B = \frac{1}{N} \left( \sum_{i=1}^{N-1} \sum_{j=i+1}^{N} RF(T_i, T_j) \right) - SS_W, \quad (4)$$

where $T_i$ and $T_j$ are trees $i$ and $j$ in a given set of trees $\Pi$, and $SS_W$ is the index of intragroup evaluation defined in Equation 2.

The results of our simulation study suggest that this version of *CH* index outperforms the popular Silhouette (*SH*), Ball and Hall (*BH*), and Gap cluster validity indices adapted for tree clustering with $k$-means (see Fig. 1, and Appendix S2 for their definitions). Moreover, the version of our algorithm based on the objective function $OF_{EA}$ and *CH* index generally outperforms the tree clustering algorithms of Stockham *et al.* (2002), Tahiri *et al.* (2018) and Bonnard *et al.* (2006) (see Fig. 2). However, in the case of large trees, it provides the results equivalent to those yielded by the Stockham algorithm based on computation of majority-rule consensus trees (see Fig. 2b).

## 2.2  Approximation by the lower and the upper bounds, and by their mean

As discussed above, the Euclidean objective function $OF_{EA}$ (Equation 2) can also be viewed as an approximation of the objective function *OF* defined in Equation 1. For instance, the consensus tree of the four trees presented in Supplementary Figure S1 is a star tree, and the value of the objective function *OF* for them is $(2 + 2 + 2 + 2) = 8$, whereas the value of $OF_{EA}$ is $(2 + 2 + 4 + 4 + 2 + 2)/8 = 4.5$. Thus, it would be interesting to find the lower and the upper bounds of the objective function $OF$, and use them in the clustering process. Theorem 1 below allows us to establish these bounds.

**Theorem 1** *For a given cluster $k$ containing $N_k$ phylogenetic trees (i.e. additive trees or X-trees) the following inequalities hold:*

$$\frac{1}{N_k-1}\sum_{i=1}^{N_k-1}\sum_{j=i+1}^{N_k}RF(T_{ki},T_{kj}) \leq RF(C_k,T_{ki}) \leq \frac{2}{N_k}\sum_{i=1}^{N_k-1}\sum_{j=i+1}^{N_k}RF(T_{ki},T_{kj}),\quad(5)$$

*where $N_k$ is the number of trees in cluster $k$, $T_{ki}$ and $T_{kj}$ are, respectively, trees $i$ and $j$ in cluster $k$, and $C_k$ is the majority-rule consensus tree of cluster $k$.*

The proof of Theorem 1 is presented in Appendix S3 (see Supplementary Material). It is worth noting that the lower and the upper bounds of $\sum_{i=1}^{N_k}RF(C_k,T_{ki})$ defined in Theorem 1 are identical when $N_i = 2$. Moreover, the value of the objective function defined by the upper bound of (5) divided by 2 (i.e. the function $OF_{EA}$ defined in Equation 2) is smaller than the value of the objective function $OF$ (Equation 1). Thus, according to the terminology of approximation theory, the criterion $OF_{EA}$ is a factor-2 approximation of the criterion $OF$.

We can use these bounds as well as the middle of the interval defined by them, which is $\frac{3N_k-2}{2N_k(N_k-1)}\sum_{i=1}^{N_k-1}\sum_{j=i+1}^{N_k}RF(T_{ki},T_{kj})$, as an approximation of the contribution of cluster $k$ to the $k$-means objective function $OF$ defined by Equation 1.

For instance, the following objective function based on the middle of the interval established in Theorem 1 can also be used as an approximation of the original objective function $OF$:

$$OF_{MA} = \sum_{k=1}^{K}\frac{3N_k-2}{2N_k(N_k-1)}\sum_{i=1}^{N_k-1}\sum_{j=i+1}^{N_k}RF(T_{ki},T_{kj}).\quad(6)$$

In a similar way, we can define the objective function based on the lower bound of $OF$:

$$OF_{LA} = \sum_{k=1}^{K}\frac{1}{N_k-1}\sum_{i=1}^{N_k-1}\sum_{j=i+1}^{N_k}RF(T_{ki},T_{kj}),\quad(7)$$

and the upper bounds of $OF$ as well:

$$OF_{UA} = \sum_{k=1}^{K}\frac{2}{N_k}\sum_{i=1}^{N_k-1}\sum_{j=i+1}^{N_k}RF(T_{ki},T_{kj}).\quad(8)$$

Clearly, the use of the approximation functions defined by Equations (6-8) does not increase the time complexity of our clustering method, which will remain $O(nN^2 + rKNI)$. The use of these approximation functions should imply the appropriate changes in the formulas of the considered cluster validity indices (see Equations 21-22 in Supplementary Material).

Importantly, the Euclidean objective function $OF_{EA}$ and the upper bound objective function $OF_{UA}$ defined in Equations 2 and 8, respectively, differ by a constant multiplier only. This means that they reach their minimum at the same point (i.e. the same tree clustering). Thus, only one of these functions (i.e. $OF_{EA}$) was tested in our simulations along with $OF_{LA}$ and $OF_{MA}$.

### 2.3 Clustering trees with different sets of leaves - the supertree approach

In this section, we explain how the tree clustering method introduced above for the case of trees defined on the same set of leaves could be extended to trees whose sets of leaves can differ, as it is often the case in phylogenetic studies, including the famous Tree of Life project (Maddison *et al.* 2007).

Let $\Pi$ be a set of $N$ unrooted phylogenetic trees that may contain different, but mutually overlapping, sets of labeled leaves. In this case, the original objective function $OF$ (Equation 1) can be reformulated as follows:

$$OF_{ST} = \sum_{k=1}^{K}\sum_{i=1}^{N_k}RF_{norm}(ST_k,T_{ki}) = \sum_{k=1}^{K}\sum_{i=1}^{N_k}\left(\frac{RF(ST_k,T_{ki})}{2n(ST_k,T_{ki})-6}\right),\quad(9)$$

where $K$ is the number of clusters, $N_k$ is the number of trees in cluster $k$, $RF_{norm}(ST_k,T_{ki})$ is the normalized Robinson and Foulds topological distance between tree $i$ of cluster $k$, denoted $ST_k$, and the majority-rule supertree of this cluster, denoted $ST_k$, reduced to a subtree having all leaves in common with $T_{ki}$. The reduced version of the supertree $ST_k$ is obtained after removing from it all leaves that do not belong to $T_{ki}$ and collapsing the corresponding branches. The $RF$ distance is normalized by dividing it by its maximum possible value, which is $2n(ST_k,T_{ki})$-6, where $n(ST_k,T_{ki})$ is the number of common leaves in trees $ST_k$ and $T_{ki}$. The $RF$ normalization is carried out to account equally the contribution of each tree to clustering. Obviously, Equation 9 can be considered only if the number of common leaves in $ST_k$ and $T_{ki}$ is greater than 3.

We propose to use the following analogue of the Euclidean approximation function (see Equation 2) to avoid supertree recomputations at each step of $k$-means:

$$OF_{ST\_EA} = \sum_{k=1}^{K}\frac{1}{N_k}\sum_{i=1}^{N_k-1}\sum_{j=i+1}^{N_k}\left(\frac{RF(T_{ki},T_{kj})}{2n(T_{ki},T_{kj})-6}+\alpha\times\frac{n(T_{ki})+n(T_{kj})-2n(T_{ki},T_{kj})}{n(T_{ki})+n(T_{kj})}\right),\quad(10)$$

where $n(T_{ki})$ is the number of leaves in tree $T_{ki}$, $n(T_{kj})$ is the number of leaves in tree $T_{kj}$, $n(T_{ki},T_{kj})$ is the number of common leaves in trees $T_{ki}$ and $T_{kj}$, and $\alpha$ is the penalization (tuning) parameter, taking values between 0 and 1, used to prevent from putting to the same cluster trees having small percentages of leaves in common. This penalization parameter is necessary to get well-balanced clusters in which trees have both high topological and species content similarity. Indeed, the normalized $RF$ distance between two large trees can be small only because the trees do not have enough taxa in common. However, such trees should not be necessarily assigned to the same cluster. Equation 10 also implies that two trees belonging to the same cluster have at least four taxa in common, but a higher taxa-similarity threshold can be used as the method's parameter in order to increase the cluster homogeneity. The objective functions reported in Equations (6-8) and the corresponding cluster validity indices should be normalized in a similar way.

In case of supertree clustering, the $SS_W$ index can be computed as follows:

$$SS_W = \sum_{k=1}^{K}\frac{1}{N_k}\sum_{i=1}^{N_k-1}\sum_{j=i+1}^{N_k}\left(\frac{RF(T_{ki},T_{kj})}{2n(T_{ki},T_{kj})-6}+\alpha\times\frac{n(T_{ki})+n(T_{kj})-2n(T_{ki},T_{kj})}{n(T_{ki})+n(T_{kj})}\right),\quad(11)$$

and the $SS_B$ index as follows:

$$SS_B = \frac{1}{N}\left(\sum_{i=1}^{N-1}\sum_{j=i+1}^{N}\left(\frac{RF(T_i,T_j)}{2n(T_i,T_j)-6}+\alpha\times\frac{n(T_i)+n(T_j)-2n(T_i,T_j)}{n(T_i)+n(T_j)}\right)\right)-SS_W.\quad(12)$$

The clustering procedure based on the use of Equation 10 should be carried out with different random input partitions. The best clustering solution can be selected using the value of the adapted $CH$ validity index based on Equations (11-12) and Equations (15-17) in Supplementary Material. Once the best clusters of trees are chosen, any existing supertree reconstruction method (Bininda-Emonds 2004) can be applied to infer the related majority-rule supertrees.

## 3  Results

In this section, we first present our simulation design (sub-section 3.1) and then discuss the results of our simulation study conducted with synthetic data (sub-section 3.2). Finally, in sub-section 3.3, we describe the results of our clustering analysis of the SARS-CoV-2 data, originally examined by Lam *et al.* (2020) and Makarenkov *et al.* (2021).

### 3.1  Simulation design

We tested our new method for computing multiple consensus trees and supertrees using a protocol that included three different simulation experiments. The first experiment involved partitioning phylogenies having identical sets of leaves (i.e. multiple consensus trees were constructed) and included a comparison of our new method with some state-of-the-art tree partitioning algorithms. The second experiment consisted in partitioning phylogenies having different sets of leaves (i.e. multiple consensus supertrees were constructed) without using penalization in the objective function (i.e. the value of the penalization parameter $\alpha$ in Equation 10 was set to 0). The third experiment involved partitioning phylogenies having different sets of leaves using the penalization parameter $\alpha$ (its value varied between 0 and 1) in the objective function (10). For reasons of both fairness and efficiency, the three tree partitioning algorithms compared in our simulations, i.e. the Stockham *et al.* (2002) algorithm, the *k*-medoids-based tree clustering algorithm by Tahiti et al. (2018), and our proposed *k*-means-based clustering algorithm, were carried out with 100 random input partitions for each dataset considered. As we could observe in our simulations, a higher number of input trees ($N$) required a higher number of random input partitions ($r$) to obtain quality results. The detailed simulation protocol adopted in our study is presented in Appendix S4 (see Supplementary Material).

## 3.2 Simulation results

The results of the simulations conducted with synthetic data are illustrated in Figures 1 and 2 (clustering of gene trees defined on the same set of taxa), Supplementary Figures S3 and S4 (see Supplementary Material), and Figure 3 (clustering of gene trees defined on different, but mutually overlapping, sets of taxa). Figure 1 presents the clustering performances of the six compared variants of our partitioning algorithm (see Appendix S4 in Supplementary Material and the legend of Fig. 1). The best overall results in this simulation were provided by the variant of our algorithm based on the $OF_{EA}$ objective function (Equation 2) and $CH$ cluster validity index (*CH-E*). The results obtained with the variant based on the $OF_{MA}$ objective function with approximation by the mean of interval (Equation 6) and $CH$ cluster validity index (*CH-MI*) were only slightly worse. At the same time, the average *ARIs* (Adjusted Rand Indices) calculated for the variants using the $BH$ and $Gap$ statistics were generally much lower than those provided by the variants based on the $CH$ and Silhouette cluster validity indices. Among the strategies that were able to deal with homogeneous data (when the number of clusters $K$ was equal to 1), the variant based on $BH$ outperformed that based on Gap for the lower numbers of clusters ($K = 1$, 2 and 3), but was less effective than it for the higher numbers of clusters ($K = 4$ and 5). It is worth noting that the results provided by the variants based on the $CH$ index (*CH-E*, *CH-MI* and *CH-LB*) were very stable; they do not vary a lot depending on the number of clusters, the number of leaves, or the coalescence rate.

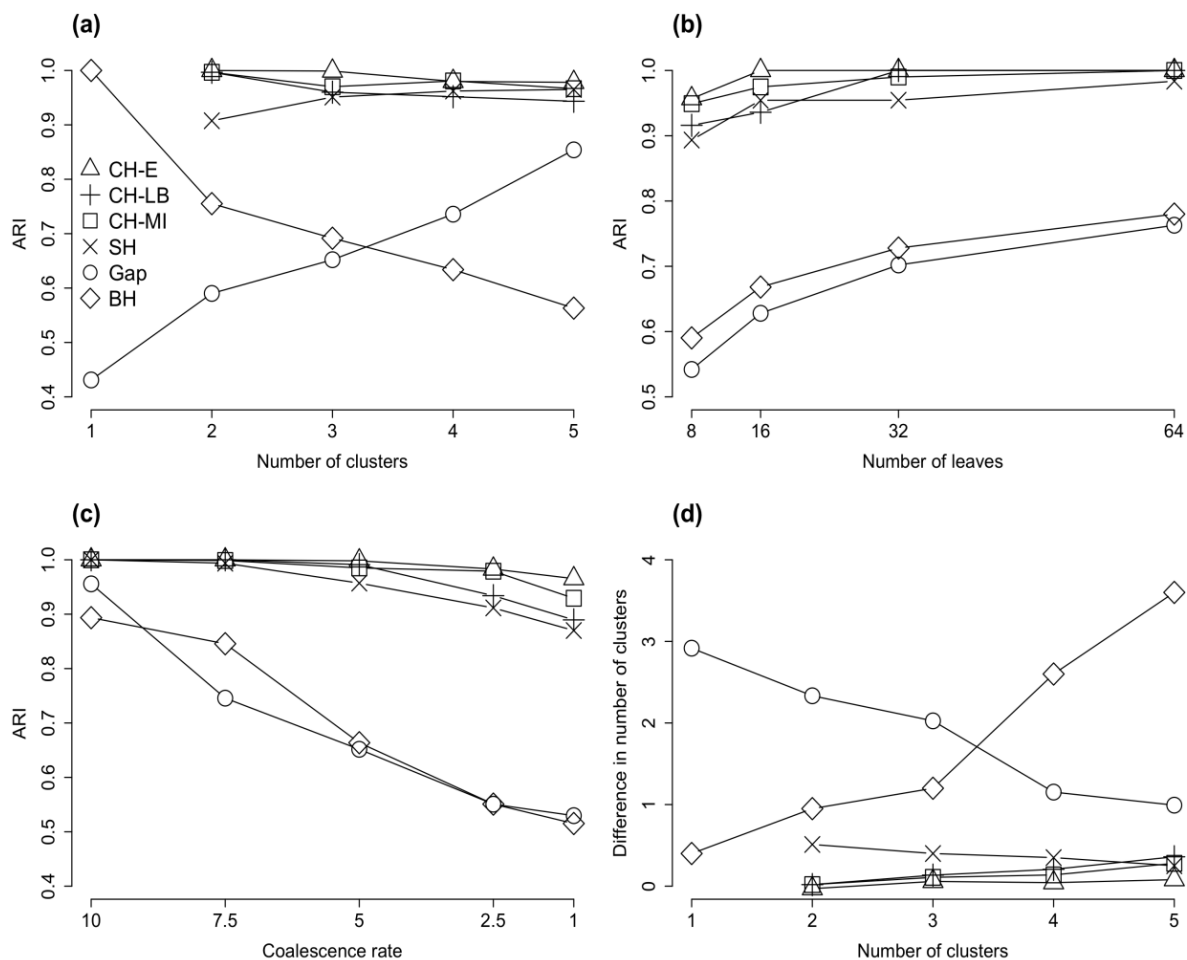

**Fig. 1.** Classification performance of the six variants of our *k*-means tree clustering algorithm applied to trees with identical sets of leaves, using respectively: $OF_{EA}$ objective function and $CH$ cluster validity index (*CH-E*), $OF_{LA}$ objective function with approximation by the lower bound and $CH$ cluster validity index (*CH-LB*), $OF_{MA}$ objective function with approximation by the mean of interval and $CH$ cluster validity index (*CH-MI*), $OF_{EA}$ objective function and Silhouette cluster validity index (*SH*), $OF_{EA}$ objective function and Gap cluster validity index (*Gap*), and $OF_{EA}$ objective function and Ball and Hall cluster validity index (*BH*). Only the $Gap$ and $BH$ indices could be used to assess the algorithm's performance on datasets containing one cluster. The results are presented in terms of the average *ARI* with respect to the: (a) number of tree clusters, (b) number of tree leaves and (c) coalescence rate, and in terms of the: (d) average absolute difference between the true and the obtained number of clusters. The coalescence rate parameter in the *HybridSim* program (Woodhams *et al.*, 2016) was set to 5 in simulations (a), (b) and (d). The presented results are the averages taken over all considered numbers of leaves in simulations (a), (c) and (d), and all considered numbers of clusters in simulations (b) and (c).

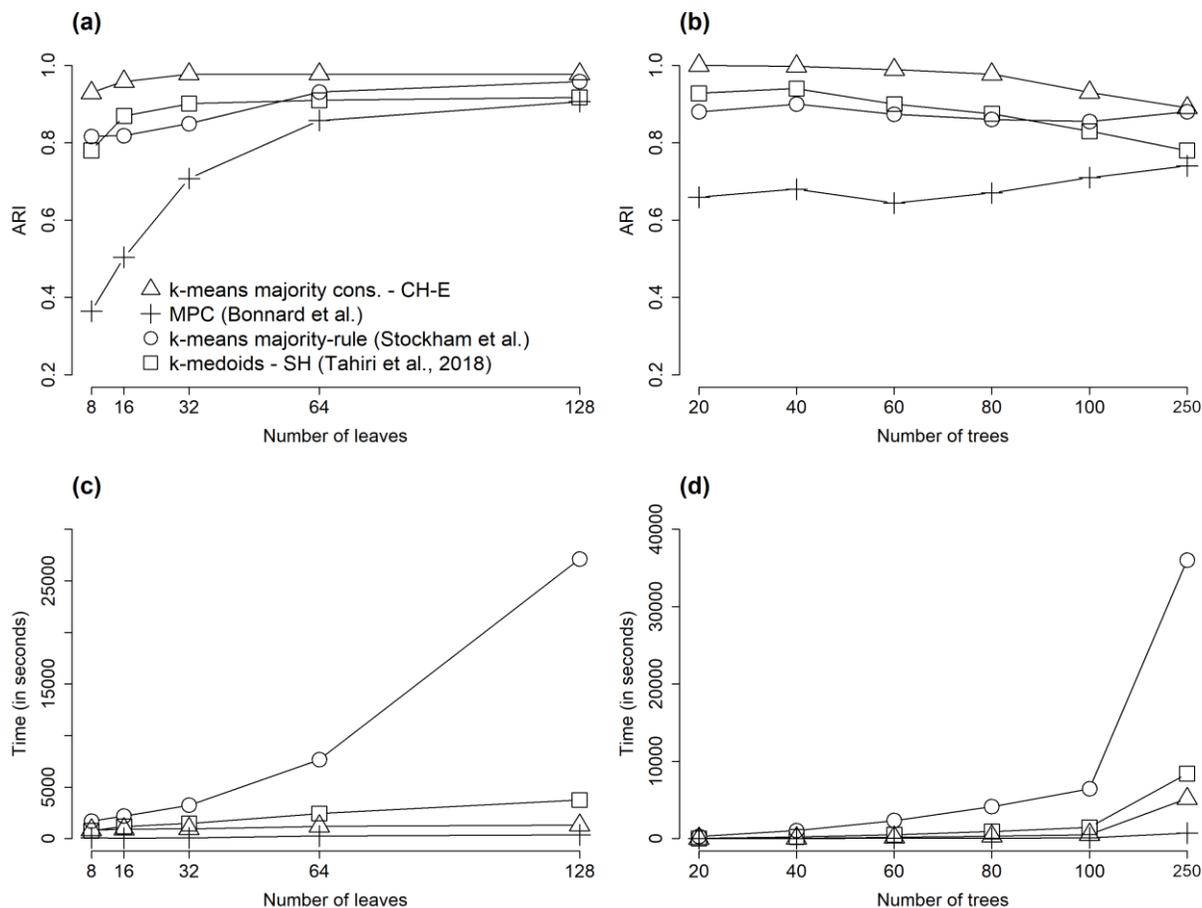

**Fig. 2.** Comparison of our algorithm (Δ) based on *k*-means tree clustering with $OF_{EA}$ objective function and *CH* cluster validity index (*CH-E*), the MPC tree clustering algorithm (+) by Bonnard *et al*. (2006), the tree clustering by Stockham *et al*. (2002) (○), and the *k*-medoids tree clustering algorithm by Tahiri *et al*. (2018) based on the *RF* distance and *SH* cluster validity index (□). The comparison was made in terms of the average *ARI* (panels a and b) and the average running time (measured in seconds) of the algorithms (panels c and d) with respect to the number of leaves and the number of trees. The coalescence rate parameter was set to 5 and the number of clusters varied from 2 to 5 in this simulation.

Figure 2 presents the results of comparison of the most stable variant of our algorithm, *CH-E*, with the state-of-the-art tree clustering methods, including the MPC tree clustering algorithm by Bonnard *et al*. (2006), the tree clustering algorithm by Stockham *et al*. (2002), which uses the squared *RF* distance and recomputes the majority-rule consensus trees at each iteration of *k*-means, and the *k*-medoids tree clustering algorithm by Tahiri *et al*. (2018) based on the *RF* distance (non-squared) and the *SH* cluster validity index. The curves presented in Figure 2 (a and b) indicate that our *CH-E* strategy clearly outperformed the three other competing methods in terms of the clustering quality (i.e. average *ARI* results). The clustering results provided by the methods of Stockham *et al*. and Tahiri *et al*. were very close, and both of them generally outperformed the MPC method. The only case where the clustering performances of the four competing methods were almost similar was the case of large phylogenies (trees with 128 leaves). All the methods, but especially MPC, showed an increase in the *ARI* values as the number of tree leaves increases. Moreover, *CH-E* and MPC were by far the best methods in terms of the running time for both simulation parameters considered: the number of tree leaves (Fig 2c) and the number of trees (Fig 2d). These results suggest that our new algorithm, along with MPC, is well suited for analyzing large phylogenetic datasets. However, for smaller phylogenies (with < 128 leaves) our *CH-E* strategy represents the best choice overall.

Figure 3 illustrates the performance of our supertree clustering algorithm applied to gene trees with different numbers and sets of leaves using the $OF_{ST\_EA}$ objective function (Equation 10) and *CH* cluster validity index. In this experiment, the value of the penalization parameter α in Equation 10 varied from 0 to 1, with the step of 0.2. Obviously, when no species are removed from the dataset, the penalization term in Equation 10 equals 0 and has no impact on the clustering process. However, in all other cases, the presented *ARI* curves showed different behaviour and reached its maxima at different values of α (i.e. for the case of 10% of missing species the maximum value of *ARI* was reached at α = 0, for the case of 25% of missing species the maximum was reached at α = 0.2, with very close *ARI* results obtained for α = 0.4 and 1.0, whereas for the case of 50% of missing species the maximum was reached at α = 1.0). Thus, we can conclude that the value of the penalization parameter α must be chosen with respect to the number of missing taxa in the input trees. Classification performances of our supertree clustering algorithm based on the $OF_{ST\_EA}$ objective function used with *CH* and *BH* cluster validity indices are presented in Appendix S5 (see Figs. S3 and S4 in Supplementary Material).

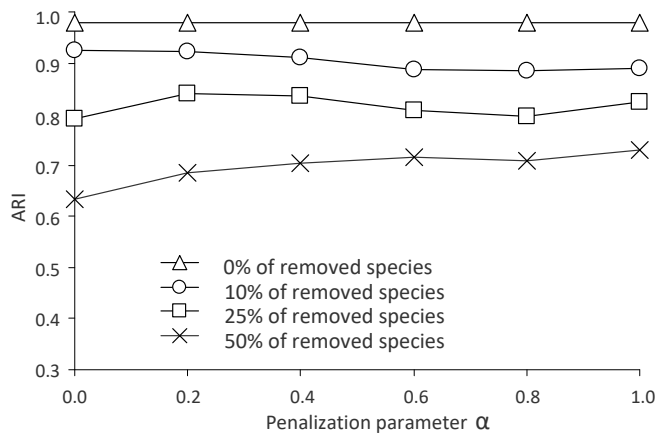

**Fig. 3.** Classification performance of our *k*-means-based supertree clustering algorithm applied to trees with different numbers and sets of leaves using the $OF_{ST\_EA}$ objective function and $CH$ cluster validity index. The results are presented in terms of *ARI* with respect to the value of the penalization parameter α, which varied between 0 and 1 with the step of 0.2. The coalescence rate parameter in the *HybridSim* program was set to 5 in this simulation. The presented results are the average ARIs computed overall numbers of leaves and clusters considered in our simulations.

### 3.3 Supertree clustering of SARS-CoV-2 gene trees

We performed the SARS-CoV-2 gene tree clustering to identify genes having similar evolutionary patterns. Our analysis was conducted using the supertree clustering algorithm described in Section 2.3. It was carried out for 11 main genes of the SARS-Cov-2 genome (i.e. genes ORF1ab, S, ORF3a, E, M, ORF6, ORF7a, ORF7b, ORF8, N and ORF10) as well as for the RD domain of the spike protein because of its key evolutionary importance. Thus, 12 gene phylogenies were inferred and analyzed in our study. It is worth mentioning that some gene annotations were absent in the GenBank and Gisaid databases (i.e. some taxonomic annotations for genes ORF6, ORF7a, ORF7b, ORF8 and ORF10 were unavailable), thus leading to gene phylogenies having different, but mutually overlapping, sets of leaves. Multiple sequence alignments for all gene and genome sequences used in this study as well as all inferred gene and species phylogenies (in the Newick format) are available at: http://www.info2.uqam.ca/~makarenkov_v/Supplementary_Material_data.zip (the complete data description is available in Appendix S6 in Supplementary Material).

In total, 7 gene trees with 43 leaves (the phylogenies of genes ORF1ab, S, ORF3a, E, M, N and that of the RB domain), 4 gene trees with 25 leaves (the phylogenies of genes ORF6, ORF7a, ORF7b and ORF10), and 1 gene tree with 23 leaves (the phylogeny of gene ORF8) were considered. The internal branches of gene phylogenies with bootstrap support lower than 50% were collapsed prior to conducting gene tree clustering. The average number of missing leaves by gene tree was 17.8%. According to our simulations (see Fig. 3), the optimal value of the penalization parameter α could vary between 0 and 0.2 for such data. Here, we present the clustering results obtained with the value of α = 0.1 (very similar clusterings were obtained with α = 0 and α = 0.2; only one tree changed its cluster membership with α = 0.2). The obtained clustering solution with 3 disjoint clusters is presented in Figure 4 (a, b and c). This solution encompasses three main patterns of evolution of betacoronavirus genes. The phylogenies of genes E, M, ORF6, ORF7a, ORF7b, ORF8 and ORF10 were assigned to Cluster 1, those of genes ORF1ab, S, ORF3a and N to Cluster 2, and that of the RB domain to Cluster 3. The consensus supertrees for each cluster were then inferred using the CLANN program (Creeve and McInerney 2005). The supertree for Cluster 1 (Fig. 4a) was inferred using the heuristic search (*hs*) and *bootstrap* (performed with 100 replicates) options available in CLANN. The supertree for Cluster 2 (Fig. S4b) was obtained using the *consensus* option of CLANN as all four trees forming this cluster contained 43 taxa. The consensus supertree of Cluster 3, containing a singleton element (i.e. the gene phylogeny of the RB domain) was its gene tree (Fig. 4c).

Moreover, we also conducted a detailed HGT (Horizontal Gene Transfer) and recombination analysis of the obtained supertrees in order to identify the main HGT and recombination patterns characterizing the evolution of SARS-CoV-2 and related betacoronaviruses. HGT and recombination are widespread reticulate evolutionary processes contributing to diversity of betacoronaviruses, as well as of most other viruses, allowing them to overcome selective pressure and adapt to new environments (Pérez-Losada et al. 2015). The HGT-Detection algorithm of Boc et al. (2010) was carried out independently for each of the three consensus supertrees inferred for our data using the T-Rex web server (Boc et al. 2012) and the Armadillo 1.1 workflow platform (Lord et al. 2012). The obtained gene transfers, representing common HGT and recombination trends of the tree cluster under study, are indicated by red arrows in Figure 4d. We then completed our analysis by conducting an independent gene transfer detection for all individual gene trees included in Clusters 1 and 2 (the detected individual HGTs that were different from the previously recovered common transfers are indicated by blue and green arrows, respectively).

The obtained clustering and HGT detection results highlight the uniqueness of the evolution of the RB domain. They also suggest that the RB domain of SARS-CoV-2 could be acquired by a horizontal transfer of genetic material, followed by intragenic recombination, from the Guangdong pangolin CoV. Furthermore, they emphasize the stability of evolutionary patterns of the longest CoV genes (i.e. ORF1ab, S, ORF3a and N assigned to Cluster 2) as the consensus supertree (i.e. consensus tree in this case) of this cluster has a well-resolved structure. The consensus supertree of Cluster 1 is less resolved of all supertrees. This could be expected since Cluster 1 contains 7 of 12 gene trees considered in our study. The topology of Cluster 1 supertree points out that SARS-CoV-2 could not only be a mosaic created by recombination of the bat RaTG13 and Guangdong pangolin CoVs, but is also a very close relative of the bat ZC45 and ZXC21 CoV strains, which are the most closely located taxa to the clade of the RaTG13 and SARS-CoV-2 viruses in this supertree.

One of the main advantages of our program is that it allows the users to process unresolved phylogenies and trees with different, but mutually overlapping, sets of leaves. It was the case of the considered SARS-CoV-2 dataset. For example, the algorithms of Stockham et al. (2002), Tahiri et al. (2018) and Bonnard et al. (2006), which work only with phylogenies defined on the same set of taxa, could be applied to this dataset only when the 12 original phylogenies were reduced to the common set of 23 taxa, thus loosing the opportunity to infer many common HGT and recombination events depicted in Fig. 4d. Moreover, for the MPC method (Bonnard et al. 2006), we needed to transform the unresolved phylogenies into binary trees. For such reduced and transformed data, the *k*-means majority rule clustering method found 4 clusters of trees, the *k*-medoids-based method generated 3 such clusters, whereas the MPC method, executed with the Welsh-Powell option, found 9 clusters. All solutions provided by these clustering methods were quite different from the clustering yielded by our method (see Fig. 4). Furthermore, none of these methods was able to emphasize the uniqueness of the RB domain evolution by putting it in a separate singleton cluster.

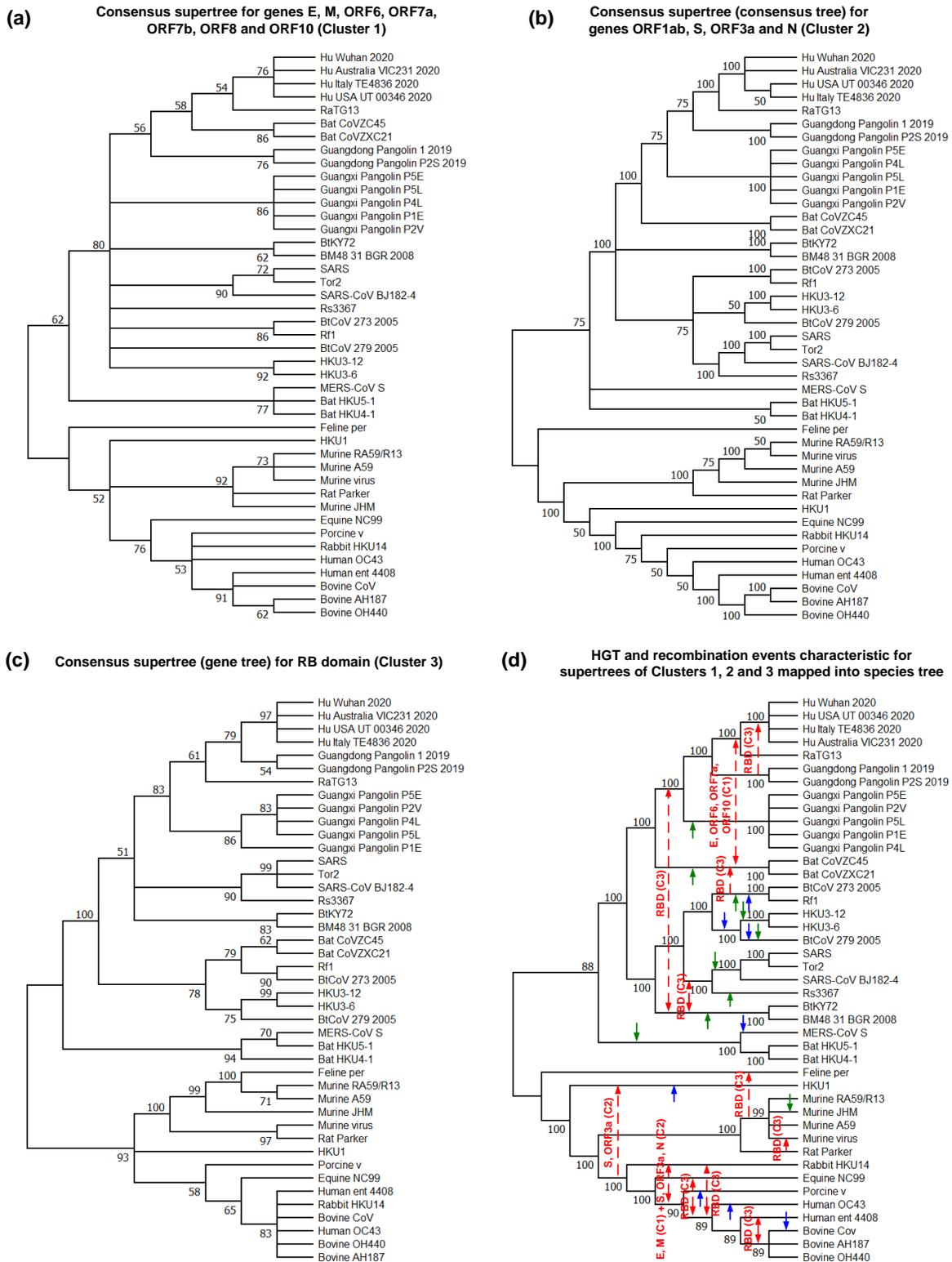

**Fig. 4.** Three consensus supertrees illustrating three main ways of evolution of betacoronavirus genes and HGT-recombination network depicting the most important gene transfer and recombination events found for each consensus supertree. (a) Supertree of Cluster 1 is the best heuristic search (*hs*) CLANN supertree inferred for the phylogenies of genes E, M, ORF6, ORF7a, ORF7b, ORF8 and ORF10; (b) Supertree of Cluster 2 is the extended majority-rule consensus tree inferred for the phylogenies of genes ORF1ab, S, ORF3a and N (defined on the same set of 43 taxa); (c) supertree of Cluster 3 is the RAxML gene tree of the RB domain of the spike protein (a unique member of this cluster); (d) Horizontal gene transfer and recombination events inferred by the HGT-Detection program for each of the three consensus supertrees. Bootstrap scores are indicated on the internal tree branches. Branches with bootstrap support lower than 50% were collapsed. Transfer directions are represented by arrows (when the direction is uncertain, the arrow is bidirectional). Gene(s) affected by the transfer and the corresponding tree cluster numbers are indicated on the red arrows. Blue (for Cluster 1) and green (for Cluster 2) arrows represent additional HGT events found by the HGT-Detection program for individual genes of these clusters (these events cannot be inferred directly from consensus supertrees).

## 4 Discussion

### 4.1 Main properties and advantages of the method

Consensus tree and supertree inference methods synthesize collections of gene phylogenies into comprehensive trees that preserve main topological features of the input phylogenies and include all taxa present in them. In this paper, we introduced a new systematic method for inferring multiple alternative consensus trees and supertrees from a given set of phylogenetic trees, which can be defined either on the same set of taxa (case of multiple consensus trees) or on different sets of taxa with incomplete taxon overlap (case of multiple supertrees). To the best of our knowledge, the problem of building multiple alternative supertrees has not been addressed yet in the literature. The inferred alternative consensus trees and supertrees represent the most important evolutionary patterns characterizing the evolution of genes under study. They are generally much better resolved than a single consensus tree or a single supertree inferred by traditional methods. Thus, a multiple consensus tree or supertree inference approach has the potential to build supertrees that retain much more plausible information from the input set of gene phylogenies. A single consensus tree or supertree could be an appropriate representation of a given set of gene trees only if all of them, or a large majority of them, follow the same evolutionary patterns. For example, our method allows one to identify ensembles of genes that underwent similar horizontal gene transfer, hybridization or intragenic/intergenic recombination events, or those that were affected by similar ancient duplication events during their evolution.

The new method relies on multiple runs of the $k$-means partitioning algorithm applied to the non-squared Robinson and Foulds distances (original or normalized) between the input trees. A number of efficient approximations of the straightforward objective function (Equation 1), preventing us from computing a consensus tree for any considered cluster of trees in the internal loop of $k$-means and using some remarkable properties of the $RF$ distance, have been introduced (see Equations 2 and 6-8). These equations allow us to precompute all $RF$ distances prior to carrying out the $k$-means tree clustering and ensure that a basic $k$-means object relocation operation, consisting in removing a given tree $T$ from its current cluster $C$ and assigning it to the best possible cluster different from $C$ (if any), can be performed in $O(K)$, where $K$ is the number of tree clusters. This property makes our method perfectly suitable for analysis of large evolutionary datasets. In order to compute the $RF$ distance between pairs of trees defined on different, but mutually overlapping, sets of leaves we propose to reduce them pairwise to common sets of leaves and then to normalize the obtained distance value. In case of supertree clustering, we also added to the objective function of the method the term including the penalization parameter $\alpha$, which is used to create well-balanced clusters that contain trees with both topological and species content similarity. This is a common way of addressing the problem of missing data in clustering, and in machine learning in general (Pan and Shen 2007).

The use of the $RF$ distance, and not of its quadratic form, in tree clustering is justified by the fact that the majority-rule consensus tree of a set of trees is a median tree of this set in the sense of the $RF$ distance (Barthélemy and McMorris 1986). Moreover, Bansal *et al.* (2010) showed that $RF$-based supertrees are supertrees that are consistent with the largest number of clades from the input trees. We also showed how the popular Caliński-Harabasz ($CH$), Silhouette ($SH$), Ball and Hall ($BH$), and $Gap$ cluster validity indices could be adapted to tree clustering with $k$-means. The $CH$ and $SH$ indices are suitable for clustering heterogeneous data (when the number of clusters $K \geq 2$), whereas the $BH$ and $Gap$ indices can be used to cluster both homogenous (when $K = 1$) and heterogeneous data. Using simulations, we demonstrated that the version of our method based on Euclidean approximation (Equation 2) typically outperforms the existing methods for building multiple alternative consensus trees, such as MPC (Bonnard *et al.* 2006), $k$-means tree clustering with squared $RF$ distance by Stockham *et al.* (2002), and $k$-medoids tree clustering by Tahiri *et al.* (2018), in terms of both clustering quality and running time.

### 4.2 Future extensions of the method

The statistical robustness of phylogenetic trees is a very important factor that should not be neglected when inferring and interpreting the results of phylogenetic analysis. We know that the $RF$ distance is twice the number of bipartitions present in one of the trees and absent in the other (Robinson and Foulds 1981). Thus, we can incorporate bootstrap scores of the input trees in the computation by considering the following objective function in the framework of consensus tree clustering:

$$OF_{EAB} = \sum_{k=1}^{K} \frac{1}{N_k} \sum_{i=1}^{N_k-1} \sum_{j=i+1}^{N_k} \left( RF(T_{ki}, T_{kj}) + \beta \times \frac{\sum_{all\ b \in CB(T_{ki}, T_{kj})} 2|bs_b(T_{ki}) - bs_b(T_{kj})|}{100} \right), (13)$$

where $bs_b(T_{ki})$ and $bs_b(T_{kj})$ are the bootstrap scores, expressed in percentages, of branch branch $b$ that induces a common bipartition in trees $T_{ki}$ and $T_{kj}$, $CB(T_{ki}, T_{kj})$ is the set of all common internal branches (i.e. branches inducing the same bipartitions) in $T_{ki}$ and $T_{kj}$, and $\beta$ is the penalization parameter, taking values between 0 and 1, used to penalize pairs of input trees with different bootstrap scores of their internal branches inducing common bipartitions. Based on Equation 13, trees with similar bootstrap support of their internal branches inducing the same bipartitions of taxa will have a greater potential to be assigned to the same cluster. It is worth mentioning that branch lengths of compared branches can be taken into account in a similar way in the objective function of the clustering algorithm. In this case, the absolute difference between bootstrap scores of the two compared branches inducing the same bipartition in $T_{ki}$ and $T_{kj}$ (see Equation 13) can be replaced by the absolute difference between their lengths, whereas the maximum of the two branch lengths will replace 100% in the fraction denominator.

Moreover, we can use the following analogue of the Euclidean approximation function (see Equation 10) to take into account bootstrap support of the internal tree branches and avoid supertree computations at each step of the supertree $k$-means clustering:

$$OF_{ST\_EAB} = \sum_{k=1}^{K} \frac{1}{N_k} \sum_{i=1}^{N_k-1} \sum_{j=i+1}^{N_k} \left( \frac{RF(T_{ki}, T_{kj}) + \beta \times \frac{\sum_{all\ b \in CB(T_{ki}, T_{kj})} 2|bs_b(T_{ki}) - bs_b(T_{kj})|}{100}}{2n(T_{ki}, T_{kj}) - 6} \right. $$
$$\left. + \ \alpha \times \frac{n(T_{ki}) + n(T_{kj}) - 2n(T_{ki}, T_{kj})}{n(T_{ki}) + n(T_{kj})} \right). (14)$$

Another interesting option for further investigations concerns the use of other popular tree distances in the objective function of the clustering algorithm. In this context, the two most promising of them seem to be the branch score distance (Kuhner and Felsenstein 1994, Gambette *et al.* 2016) and the quartet distance (Bryant *et al.* 2000).

The branch score distance is defined as follows. Let us consider two phylogenetic trees $T$ and $T'$ defined on the same set of $n$ taxa and the large set $(BP_1, BP_2, \ldots, BP_{NB})$ of all possible bipartitions existing for these taxa. For each tree, we can determine a large vector of nonnegative values $\mathbf{bp} = (b_1, b_2, \ldots, b_{NP})$, in which $b_i$ is equal to the branch length of the branch corresponding to bipartition $BP_i$ if this bipartition exists in the tree, otherwise it is equal to 0. For two trees $T$ and $T'$, whose bipartition vectors are $\mathbf{bp}$ and $\mathbf{bp'}$, the branch score distance is defined as the squared Euclidean distance between these vectors: $BSD(T, T') = \sum_{i=1}^{NB} (b_i - b'_i)^2$, where $NB$ is the number of all possible existing bipartitions.

The quartet distance ($QD$) is defined as the number of quartets of tree leaves that induce a subtree topology that occur in only one of the two compared trees. This distance has been extensively used in computational biology not only for inferring phylogenetic trees and networks (Berry *et al.* 1999, Gambette *et al.* 2012), but also for building supertrees (Snir and Rao 2008).

According to its definition, the quartet distance is a symmetric difference distance. Thus, its square root ($QD^{1/2}$) is Euclidean (Critchley and Fichet 1994) and an analogue of Equation 2, in which $RF$ is replaced by $QD$, can be used in the objective function of the clustering algorithm. An advantage of the quartet distance is that it can be computed in $O(n\log n)$, where $n$ is the number of leaves in both trees involved in computation.

Another tree distance which could be suitable for tree clustering is the Billera–Holmes–Vogtmann (*BHV*) distance (Billera *et al.* 2001). The BHV distance between weighted trees is defined as the geodesic, or shortest path, distance inside treespace in which trees are viewed as ($2n-3$)-dimensional vectors of their bipartition weights within the larger ($2^{n-1}-1$)-dimensional space of all graphs. The *BHV* distance between two trees can be computed in $O(n^4)$ and approximated in linear time (St. John 2017). The continuous treespace introduced by Billera *et al.* (2001) provides a perfect environment for computation of average trees, while the classical Euclidean mean, when applied to tree vectors, can yield vectors not corresponding to trees. In the *BHV* framework, the majority consensus tree (McMorris *et al.* 1983) corresponds to the mode and the Fréchet mean corresponds to the average of a given set of trees. The Fréchet mean tree of a given set of $N$ trees is the tree that minimizes the sum of the squared BHV distances to the given set $\Pi = \{T_1, T_2, \ldots, T_N\}$ of $N$ phylogenetic trees defined on the same set of taxa, i.e. $min_{all\ T} \sum_{i=1}^{N} BHV(T_i, T)^2$. A normalized sum of such minimum values computed independently for each considered cluster of trees can constitute an objective function of a tree clustering method based on the *BHV* distance. Fortunately, the Fréchet mean tree is unique and its approximation can be calculated in polynomial time by an iterative algorithm. However, the Fréchet mean may also demonstrate a non-Euclidean sticky behaviour, as changing an input tree does not necessarily change the mean tree of the dataset, in contrast to Euclidean space (St. John 2017). Alternatively, the median (which is a more robust estimator than the mean) of a set of trees can also be considered when clustering tree. The *BHV* distance median of a set of trees is the tree that minimizes the sum of non-squared *BHV* distances to those trees.

Another widely-used criterion for assessing differences between trees is the Subtree Prune-and-Regraft (*SPR*) distance (Hein *et al.* 1996). This distance is integrated in a variety of tree building methods exploring different tree topologies (Gascuel 2005). Moreover, Whidden *et al.* (2014) determined that the *SPR* distance can be used to build supertrees and that *SPR*-based supertrees are significantly more similar to the known species history than *RF*-based supertrees given biologically plausible rates of simulated horizontal gene transfers. The problem of computing the *SPR* distance between two trees is NP-hard (Bordewich and Semple 2005). However, in practice, its approximation can be computed using a fixed-parameter-bounded search tree algorithm in combination with a linear-time formulation of Linz and Semple's cluster reduction to solve an equivalent maximum agreement forest problem (Whidden *et al.* 2014). Nevertheless, the *SPR* distance, as well as the other popular tree topology rearrangement distances such as Nearest Neighbor Interchange (*NNI*) and Tree Bisection and Reconnection (*TBR*), has no Euclidean properties and new formulas and algorithms should be designed in order to adapt it to tree clustering.

## Acknowledgements

We would like to thank Dr Nicolas Lartillot for providing us with the MPC tree clustering program, Drs Pierre Legendre and Bogdan Mazoure for helping us with the analysis of SARS-CoV-2 data, and Dr Russell Schwartz and two anonymous reviewers of our manuscript for their helpful comments and suggestions. We also thank Compute Canada for providing access to high-performance computing facilities.

## Funding

This work was supported by Natural Sciences and Engineering Research Council of Canada, Fonds de Recherche sur la Santé du Québec and Fonds de Recherche sur la Nature et Technologies du Québec.

*Conflict of Interest:* none declared.

## References

Ball GH and Hall DJ. (1967). A clustering technique for summarizing multivariate data. *Behavioral Sciences*, **12**, 153-155.

Bansal MS *et al.* (2010). Robinson-Foulds supertrees. *Algorithms for Molecular Biology*, **5**, 1-12.

Bapteste E *et al.* (2004). Phylogenetic reconstruction and lateral gene transfer. *Trends Microbiology*, **12**, 406-411.

Barthélemy JP and McMorris FR. (1986). The median procedure for n-trees. *Journal of Classification*, **3**, 329-334.

Baum BR. (1992). Combining trees as a way of combining data sets for phylogenetic inference, and the desirability of combining gene trees. *Taxon*, **41**, 3-10.

Berry V. *et al.* (1999). Quartet cleaning: Improved algorithms and simulations. In: Algorithms - ESA'99. Lecture Notes in Computer Science, vol. 1643, Springer, Berlin, Heidelberg.

Billera LJ. *et al.* (2001). Geometry of the space of phylogenetic trees. *Advances in Applied Mathematics*, **27**, 733-767.

Bininda-Emonds OR., editor (2004). *Phylogenetic supertrees: combining information to reveal the tree of life*. Springer Science & Business Media.

Boc A. *et al.* (2010). Inferring and validating horizontal gene transfer events using bipartition dissimilarity. *Systematic Biology*, **59**, 195-211.

Boc A. *et al.* (2012). T-REX: a web server for inferring, validating and visualizing phylogenetic trees and networks. *Nucleic Acids Research*, **40(W1)**, W573-W579.

Bonnard C. *et al.* (2006). Multipolar consensus for phylogenetic trees. *Systematic Biology*, **55**, 837–43.

Bordewich M and Semple C. (2005). On the computational complexity of the rooted subtree prune and regraft distance. *Annals of Combinatorics*, **8**, 409-423.

Bryant D. *et al.* (2000). Computing the quartet distance between evolutionary trees. Proc. 11th Annual ACM-SIAM SODA. *Journal of the Society for Industrial and Applied Mathematics*, **9**, 285-286.

Bryant D. (2003). A classification of consensus methods for phylogenetics. *DIMACS series in discrete mathematics and theoretical computer science*, **61**, 163-84.

Buneman P. (1971). The recovery of trees from measures of dissimilarity. *Mathematics and the archeological and historical sciences*. Edinburgh: Edinburgh University Press. p. 387–395.

Caliński T and Harabasz J. (1974). A dendrite method for cluster analysis. *Communications Statistics Theory Methods*, **3**, 1-27.

Creevey CJ and McInerney JO. (2005). Clann: investigating phylogenetic information through supertree analyses. *Bioinformatics*, **21**, 390-392.

Critchley F and Fichet B. (1994). The partial order by inclusion of the principal classes of dissimilarity on a finite set, and some of their basic properties. In: *Classification and dissimilarity analysis*. Springer, New York, NY, 5-65.

de Queiroz A and Gatesy J. (2007). The supermatrix approach to systematics. *Trends in Ecology and Evolution*, **22**, 4-41.

Gambette P. *et al.* (2012). Quartets and unrooted phylogenetic networks. *Journal of bioinformatics and computational biology*, **10**, 1250004.

Gambette P. *et al.* (2016). Do branch lengths help to locate a tree in a phylogenetic network? *Bulletin of Mathematical Biology*, **78**, 1773-1795.

Gascuel O. (2005). *Mathematics of Evolution and Phylogeny*. Oxford (UK): Oxford University Press, 121-142.

Guénoche A. (2013). Multiple consensus trees: a method to separate divergent genes. *BMC Bioinformatics*, **14**, 46.

Hein J *et al.* (1996). On the complexity of comparing evolutionary trees. *Discrete Applied Mathematics*, **71**, 153-169.

Jansson J. *et al.* (2013). An optimal algorithm for building the majority-rule consensus tree. In: *Annual International Conference on Research in Computational Molecular Biology*. Springer, Berlin, Heidelberg, 88-99.

Kuhner MK and Felsenstein J. (1994). A simulation comparison of phylogeny algorithms under equal and unequal evolutionary rates. *Mol. Biol. Evol.*, **11**, 459-468.

Lam TTY. *et al.* (2020). Identifying SARS-CoV-2 related coronaviruses in Malayan pangolins. *Nature*, **583**, 282-285.

Lord E. *et al.* (2012). Armadillo 1.1: an original workflow platform for designing and conducting phylogenetic analysis and simulations, *PloS One*, **7(1)**:e29903.

MacQueen J. (1967). Some methods for classification and analysis of multivariate observations. In: *Proceedings of the fifth Berkeley symposium on mathematical statistics and probability*, **1**, 281-297.

Maddison DR. (1991). The discovery and importance of multiple islands of most-parsimonious trees. *Systematic Biology*, **40**, 315-328.

Maddison DR. *et al.* (2007). The tree of life web project. *Zootaxa*, **1668**, 19-40.

Mahajan M. *et al.* (2009). The planar k-means problem is NP-hard. *Lecture Notes Computer Science*, **5431**, 274-285.

Makarenkov V. *et al.* (2021). Horizontal gene transfer and recombination analysis of SARS-CoV-2 genes helps discover its close relatives and shed light on its origin. *BMC Ecology and Evolution*, **21**, 1-18.

Makarenkov V. and Leclerc B. (2000). Comparison of additive trees using circular orders. *Journal of Computational Biology*, **7**, 731-744.

McMorris FR. *et al.* (1983). A view of some consensus methods for trees. In: Numerical Taxonomy. *Proc. NATO Advanced Study Institute on Numerical Taxonomy*. Berlin, Springer Verlag.

McMorris FR and Wilkinson M. (2011). Conservative supertrees. *Systematic Biology*, **60**, 232-238.

Pan W and Shen X. (2007). Penalized model-based clustering with application to variable selection. *Journal of Machine Learning Research*, **8**, 1145−1164.

Pérez-Losada M. et al. (2015). Recombination in viruses: mechanisms, methods of study, and evolutionary consequences. *Infect. Genet. Evol.*, **30**, 296-307.

Ragan MA. (1992). Phylogenetic inference based on matrix representation of trees. *Molecular Phylogenetics and Evolution*, **1**, 53-58.

Robinson DF and Foulds LR. (1981). Comparison of phylogenetic trees. *Mathematical Biosciences*, **53**, 131-147.

Rousseeuw PJ. (1987). Silhouettes: a graphical aid to the interpretation and validation of cluster analysis. *Journal of Computational and Applied Mathematics*, **20**, 53-65.

Sevillya G. et al. (2020). Detecting horizontal gene transfer: a probabilistic approach. *BMC Genomics*, **21**, 106.

Silva AS and Wilkinson M. (2021). On defining and finding islands of trees and mitigating large island bias. *Systematic Biology*, **70**, 1282–1294.

Snir S and Rao S. (2008). Quartets MaxCut: a divide and conquer quartets algorithm. *IEEE/ACM Transactions on Computational Biology and Bioinformatics*, **7**,704-18.

St. John K. (2017). The shape of phylogenetic treespace. *Systematic Biology*, **66**, e83-e94.

Steinley D and Brusco MJ. (2007). Initializing k-means batch clustering: A critical evaluation of several techniques. *Journal of Classification*, **24**, 99-121.

Stockham C. *et al.* (2002). Statistically based postprocessing of phylogenetic analysis by clustering. *Bioinformatics*, **18**, S285-S293.

Sul SJ and Williams TL. (2008). An Experimental Analysis of Robinson-Foulds Distance Matrix Algorithms. In: *Esa*, 793-804.

Tahiri N *et al.* (2018). A new fast method for inferring multiple consensus trees using k-medoids. *BMC Evolutionary Biology*, **18**, 48.

Tibshirani R. *et al.* (2001). Estimating the number of clusters in a data set via the gap statistic. *Journal of the Royal Statistical Society B*, **63**, 411-423.

Wareham HT. (1985). An efficient algorithm for computing MI consensus trees. *B.Sc. Honours thesis*, Memorial University of Newfoundland, Canada.

Warnow T. (2018). Supertree construction: opportunities and challenges. arXiv:1805.03530v1.

Whidden C. et al. (2014). Supertrees based on the subtree prune-and-regraft distance. *Systematic Biology*, **63**, 566-581.

Wilkinson M. *et al.* (2007). Properties of supertree methods in the consensus setting. *Systematic Biology*, **56**, 330-337.

Woodhams MD. *et al.* (2016). Simulating and summarizing sources of gene tree incongruence. *Genome Biology and Evolution*, **8**, 1299-1315.

# Supplementary Material

**Appendix S1. RF is not a Euclidean distance, but its square root is Euclidean.**

The following counter-example, involving four trees $T_1$, $T_2$, $T_3$ and $T_4$ with five leaves each (see Supplementary Fig. S1), can be used to show that $RF$ is not Euclidean. It is the simplest example possible because for the trees with four leaves, the $RF$ distance has the Euclidean properties.

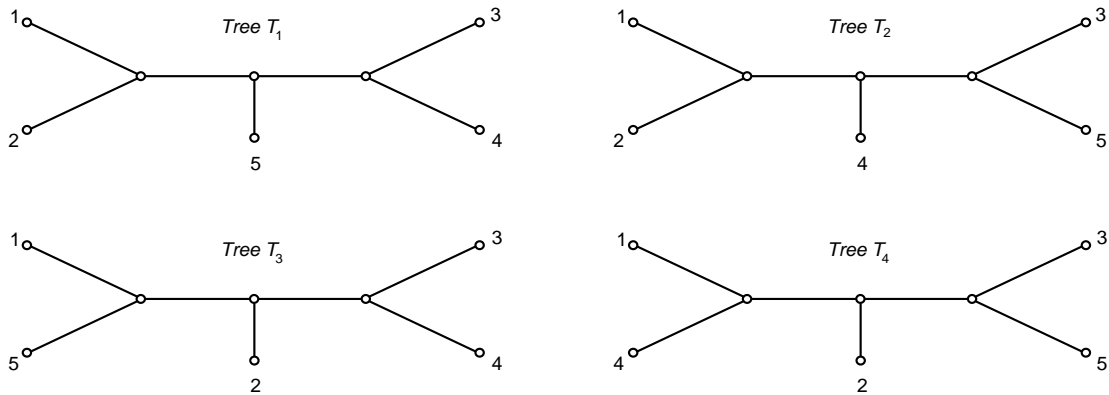

**Supplementary Fig. S1.** Four unrooted phylogenetic trees, $T_1$, $T_2$, $T_3$ and $T_4$ with five leaves used to show that the Robinson and Foulds topological distance is not a Euclidean distance.

The 1-bipartitions, corresponding to tree leaves, are common to the four trees in Supplementary Figure S1. The 2-bipartitions here, defined by the subsets of two taxa, are as follows: $T_1$: $\{1, 2\}$ and $\{3, 4\}$, $T_2$: $\{1, 2\}$ and $\{3, 5\}$, $T_3$: $\{1, 5\}$ and $\{3, 4\}$, and $T_4$: $\{1, 4\}$ and $\{3, 5\}$. Therefore, $RF(T_1, T_2) = 2$, $RF(T_2, T_3) = 4$ and $RF(T_1, T_3) = 2$. In a Euclidean case, this would place the tree $T_1$ in the middle of the interval $[T_2, T_3]$ (see Supplementary Fig. S2). At the same time, we know that $RF(T_1, T_4) = 4$ and $RF(T_3, T_4) = 4$, meaning that $T_4$ should be located on the perpendicular bisector of the interval $[T_1, T_3]$. However, this contradicts the fact that $RF(T_2, T_4) = 2$.

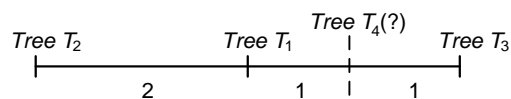

**Supplementary Fig. S2.** An illustration depicting the position of the four trees from Supplementary Figure S1 used to show that the Robinson and Foulds distance is not a Euclidean distance.

We will now recall a few mathematical results which will be useful in the sequel. They will allow us to suggest a new clustering strategy via the square root of the $RF$ distance (see the main text). In a series of beautiful papers, Kelly and Deza introduced hypermetric spaces and then extended them to quasi-hypermetric spaces (Kelly 1972; Deza and Laurent 1997). Quasi-hypermetric spaces satisfy a so-called inequality of negative type, i.e. given a metric space ($X$, $d$): for every strictly positive integer $k$, for every real numbers $\lambda_1, ..., \lambda_k$, and for all variables $x_1, ..., x_k$ in $X$, the following inequality holds: $\sum_{i=1,k} \sum_{j=1,k} \lambda_i \lambda_j d(x_i, x_j) \leq 0$. Then, it has been observed that this inequality corresponds exactly to the Schoenberg condition for the square root ($d^{1/2}$) of $d$ being of the Euclidean type. Kelly (1972) provided several examples of quasi-hypermetric spaces, such as normed spaces and lattices, and established that a symmetric difference distance is hypermetric. Therefore, its square root ($d^{1/2}$) is Euclidean (see Critchley and Fichet (1994) for more details). Because the Robinson and Foulds distance is a symmetric difference distance (Robinson and Foulds 1981), its square root ($RF^{1/2}$) is Euclidean.

## Appendix S2. Caliński-Harabasz, Silhouette, Gap and Ball-Hall cluster validity indices adapted for tree clustering with $k$-means.

### *Caliński-Harabasz cluster validity index adapted for tree clustering with k-means*

The first cluster validity index we consider here is the Caliński-Harabasz index (Caliński and Harabasz 1974). This index, sometimes called the variance ratio criterion, is defined as follows:

$$CH = \frac{SS_B}{SS_W} \times \frac{N-K}{K-1},$$ (15)

where $SS_B$ is the index of intergroup evaluation, $SS_W$ is the index of intragroup evaluation, $K$ is the number of clusters and $N$ is the number of objects (i.e. trees in our case). The optimal number of clusters corresponds to the greatest value of $CH$.

In the traditional version of $CH$, when the Euclidean distance is considered, the $SS_B$ coefficient is evaluated by using the $L^2$-norm:

$$SS_B = \sum_{k=1}^{K} N_k \left\| m_k - m \right\|^2,$$ (16)

where $m_k$ ($k = 1 ... K$) is the centroid of cluster $k$, $m$ is the overall mean (i.e. centroid) of all objects in the given dataset $X$ and $N_k$ is the number of objects in cluster $k$. In the context of the

Euclidean distance, the $SS_W$ index can be calculated using the two following equivalent expressions:

$$SS_W = \sum_{k=1}^{K} \sum_{i=1}^{N_k} \left\| x_{ki} - m_k \right\|^2 = \sum_{k=1}^{K} \frac{1}{N_k} \left( \sum_{i=1}^{N_k-1} \sum_{j=i+1}^{N_k} \left\| x_{ki} - x_{kj} \right\|^2 \right), \qquad (17)$$

where $x_{ki}$ and $x_{kj}$ are the objects $i$ and $j$ of cluster $k$, respectively (Caliński and Harabasz 1974).

To use the analogues of Equations 16 and 17 in tree clustering, we need to define the concept of centroid for a given set of trees. The *median tree* (Barthélemy and Monjardet 1981; Barthélemy and McMorris 1986) plays the role of this centroid in our tree clustering algorithm. The median procedure is defined as follows (Barthélemy and Monjardet 1981). The median trees, Md($\Pi$), for a given set of trees $\Pi = \{T_1, \ldots, T_N\}$ having the same set of leaves $S$, is the set of all trees $T$ defined on $S$, such that: $\sum_{i=1}^{N} RF(T, T_i)$ is minimized. If $N$ is odd, then the *majority-rule consensus tree*, Maj($\Pi$) of $\Pi$, is the only element of Md($\Pi$). If $N$ is even, then Md($\Pi$) is composed of Maj($\Pi$) and of some more resolved trees (see Barthélemy and Monjardet 1981 for more details).

We propose to use approximation formulas based on the properties of the Euclidean distance in order to define $SS_B$ and $SS_W$ in $k$-means-like tree clustering. These formulas do not require the computation of the majority (or the extended majority)-rule consensus trees at each iteration of $k$-means. Precisely, we replace the term $\left\| x_{ki} - x_{kj} \right\|^2$ in Equation 17 by $RF(T_{ki}, T_{kj})$ in order to obtain the approximation formula for $SS_W$:

$$SS_W = \sum_{k=1}^{K} \frac{1}{N_k} \sum_{i=1}^{N_k-1} \sum_{j=i+1}^{N_k} RF(T_{ki}, T_{kj}), \qquad (18)$$

where $T_{ki}$ and $T_{kj}$ are trees $i$ and $j$ of cluster $k$, respectively.

Also, in the case of the Euclidean distance we have:

$$SS_B + SS_W = \frac{1}{N} \left( \sum_{i=1}^{N-1} \sum_{j=i+1}^{N} \left\| x_i - x_j \right\|^2 \right), \qquad (19)$$

where $x_i$ and $x_j$ are two different objects in $X$ (Caliński and Harabasz 1974).

Thus, the approximation of the global variance between groups, $SS_B$, can be calculated as follows:

$$SS_B = \frac{1}{N}(\sum_{i=1}^{N-1}\sum_{j=i+1}^{N}RF(T_i,T_j)) - SS_W,\qquad(20)$$

where $T_i$ and $T_j$ are trees $i$ and $j$ in a given set of trees $\Pi$, and $SS_W$ is calculated according to Equation 18.

As the square root of the Robinson and Foulds distance has the Euclidean properties, Equations 18 and 20 establish the exact formulas for calculating the indices $SS_B$ and $SS_W$ for the objective function $OF_{EA}$ defined by Equation 2. Obviously, the objective function $OF_{EA}$ is only an approximation of the objective function defined in Equation 1 because the centroid of a cluster of trees is not necessarily a consensus tree of the cluster. Moreover, it is not necessarily a tree. However, as we show in the Results section (see the main text), this approximation provides very good classification results when clustering trees.

The use of the approximation functions $OF_{MA}$, $OF_{LA}$ and $OF_{UA}$ should imply the appropriate changes in the formulas of the considered cluster validity indices. For instance, in case of the Caliński-Harabasz index (*CH*) and the objective function $OF_{MA}$, the intergroup evaluation index $SS_B$ and the intragroup evaluation index $SS_W$ should be calculated as follows:

$$SS_W(OF_{MA}) = \sum_{k=1}^{K}\frac{3N_k-2}{2N_k(N_k-1)}\sum_{i=1}^{N_k-1}\sum_{j=i+1}^{N_k}RF(T_{ki},T_{kj}),\text{ and}\qquad(21)$$

$$SS_B(OF_{MA}) = \frac{3N-2}{2N(N-1)}(\sum_{i=1}^{N-1}\sum_{j=i+1}^{N}RF(T_i,T_j)) - SS_W(OF_{MA}).\qquad(22)$$

### *Silhouette index adapted for tree clustering*

Another popular index considered in our study is the Silhouette width (*SH*) (Rousseeuw 1987). Traditionally, the Silhouette width of cluster $k$ is defined as follows:

$$s(k) = \frac{1}{N_k}\left[\sum_{i=1}^{N_k}\frac{b(i)-a(i)}{\max(a(i),b(i))}\right],\qquad(23)$$

where $N_k$ is the number of objects belonging to cluster $k$, $a(i)$ is the average distance between object $i$ and all other objects belonging to cluster $k$, and $b(i)$ is the smallest, over all clusters $k$' different from $k$, of all average distances between $i$ and all the objects of cluster $k$'.

We used Equations 24 and 25 for calculating $a(i)$ and $b(i)$, respectively, in our tree clustering algorithm (see also Tahiri et al. 2018). For instance, the quantity $a(i)$ can be determined as follows:

$$a(i) = \frac{\sum_{j=1}^{N_k} RF(T_{ki}, T_{kj})}{N_k},$$ (24)

and the formula for $b(i)$ is as follows:

$$b(i) = \min_{1 \leq k' \leq K, k' \neq k} \frac{\sum_{j=1}^{N_{k'}} RF(T_{ki}, T_{k'j})}{N_{k'}},$$ (25)

where $T_{k'j}$ is tree $j$ of cluster $k'$, such that $k' \neq k$, and $N_{k'}$ is the number of trees in cluster $k'$.

The optimal number of clusters, $K$, corresponds to the maximum average Silhouette width, $SH$, which is calculated as follows:

$$SH = \bar{s}(K) = \sum_{k=1}^{K} [s(k)] / K.$$ (26)

The value of the Silhouette index defined by Equation 26 is located between -1 and +1.

***Gap statistic adapted for tree clustering***

Unfortunately, the $CH$ and $SH$ cluster validity indices defined by Equations 15 and 26 do not allow us to compare the solution consisting of a single consensus tree ($K = 1$; the calculation of $CH$ and $SH$ is impossible in this case) with clustering solutions involving multiple consensus trees ($K \geq 2$). This can be viewed as an important drawback of the $CH$ and $SH$-based classifications because a good tree clustering method should be able to recover a single consensus tree when the input set of trees is homogeneous (e.g. in case of gene trees that share the same evolutionary history).

The *Gap* statistic was first used by Tibshirani et al. (2001) to estimate the number of clusters provided by partitioning algorithms. The formulas proposed by Tibshirani et al. were based on the properties of the Euclidean distance. In the context of tree clustering, the *Gap* statistic can be defined as follows. Consider a clustering of $N$ trees into $K$ non-empty clusters, where $K \geq 1$. We first define the total intracluster distance, $D_k$, characterizing the cohesion between the trees belonging to the same cluster $k$:

$$D_k = \sum_{i=1}^{N_k} \sum_{j=1}^{N_k} RF(T_{ki}, T_{kj}).\tag{27}$$

Then, the sum of the average total intracluster distances, $V_K$, can be calculated:

$$V_K = \sum_{k=1}^{K} \frac{1}{2N_k} D_k.\tag{28}$$

Finally, the *Gap* statistic, which reflects the quality of a given clustering solution with $K$ clusters, can be defined as follows:

$$Gap_N(K) = E_N^*\{\log(V_K)\} - \log(V_K),\tag{29}$$

where $E_N^*$ denotes expectation under a sample of size $N$ from the reference distribution. The following formula (Tibshirani et al. 2001) for the expectation of $\log(V_K)$ was used in our method:

$$E_N^*\{\log(V_K)\} = \log(Nn/12) - (2/n)\log(K),\tag{30}$$

where $n$ is the number of tree leaves.

The largest value of the *Gap* statistic corresponds to the best clustering.

### *Ball-Hall index adapted for tree clustering*

Ball and Hall (1965) introduced the ISODATA procedure to measure the average dispersion of groups of objects with respect to the mean square root distance, i.e. the intra-group distance, which would lead to the following formula in case of tree clustering:

$$BH = \frac{1}{K} \sum_{k=1}^{K} \frac{1}{N_k} \sum_{i=1}^{N_k} RF(C_k, T_{ki}).\tag{31}$$

Replacing the inner sum of Equation 31 by its Euclidean approximation (as in Equation 2), we obtain the following formula:

$$BH = \frac{1}{K} \sum_{k=1}^{K} \frac{1}{N_k^2} \sum_{i=1}^{N_k-1} \sum_{j=i+1}^{N_k} RF(T_{ki}, T_{kj}).\tag{32}$$

## Appendix S3. Theorem establishing the lower and the upper bounds of the objective function *OF*.

### *Theorem* 1

*For a given cluster k containing $N_k$ phylogenetic trees (i.e. additive trees or X-trees) the following inequalities hold*:

$$\frac{1}{N_k-1} \sum_{i=1}^{N_k-1} \sum_{j=i+1}^{N_k} RF(T_{ki}, T_{kj}) \le \sum_{i=1}^{N_k} RF(C_k, T_{ki}) \le \frac{2}{N_k} \sum_{i=1}^{N_k-1} \sum_{j=i+1}^{N_k} RF(T_{ki}, T_{kj}),$$

where $N_k$ is the number of trees in cluster $k$, $T_{ki}$ and $T_{kj}$ are, respectively, trees $i$ and $j$ in cluster $k$, and $C_k$ is the majority-rule consensus tree of cluster $k$.

_Proof_

First, the sum $\sum_{i=1}^{N_k} RF(C_k, T_{ki})$ can be decomposed into the following double sum:

$$\sum_{i=1}^{N_k} RF(C_k, T_{ki}) = \sum_{i=1}^{N_k-1} \sum_{j=i+1}^{N_k} \frac{1}{(N_k-1)}(RF(C_k, T_{ki}) + RF(C_k, T_{kj})). \tag{33}$$

We know that the Robinson and Foulds distance is a metric, and thus satisfies the triangular inequality (Robinson and Foulds 1981). Hence, the following inequality holds for any pair of trees $(T_{ki}, T_{kj})$: $RF(C_k, T_{ki}) + RF(C_k, T_{kj}) \geq RF(T_{ki}, T_{kj})$.

This means that:

$$\sum_{i=1}^{N_k} RF(C_k, T_{ki}) \geq \frac{1}{(N_k-1)} \sum_{i=1}^{N_k-1} \sum_{j=i+1}^{N_k} RF(T_{ki}, T_{kj}). \tag{34}$$

Second, based on the property, proved by Barthélemy and McMorris, that the majority consensus tree of a set of trees is a median tree of this set in the sense of the _RF_ distance (Barthélemy and McMorris 1986; Barthélemy and Monjardet 1981), we have: $\sum_{i=1}^{N_k} RF(C_k, T_{ki}) = \underset{T \in \mathrm{T}(n)}{Min} \sum_{i=1}^{N_k} RF(T, T_{ki})$, where $\mathrm{T}(n)$ is the set of all possible phylogenetic trees with $n$ leaves.

Thus, we obtain:

$$\sum_{i=1}^{N_k} RF(C_k, T_{ki}) \leq \underset{1 \leq j \leq N_k}{Min}(\sum_{i=1}^{N_k} RF(T_{kj}, T_{ki})) \leq \frac{1}{N_k} \sum_{i=1}^{N_k} \sum_{j=i}^{N_k} RF(T_{kj}, T_{ki}). \tag{35}$$

It is easy to see that the upper bound in Equation 35 equals to: $\frac{2}{N_k} \sum_{i=1}^{N_k-1} \sum_{j=i+1}^{N_k} RF(T_{ki}, T_{kj})$. Obviously, the term $\underset{1 \leq j \leq N_k}{Min}(\sum_{i=1}^{N_k} RF(T_{kj}, T_{ki}))$ can be also used as an upper bound of $\sum_{i=1}^{N_k} RF(C_k, T_{ki})$. □

**Appendix S4. Detailed simulation protocol.**

This section presents the detailed simulation protocol adopted in our study. The data generation procedure used in the first experiment (i.e. with multiple consensus trees) included three main steps. First, we randomly generated a species phylogeny $T_0$ with $n$ leaves (i.e. it played the role of the first consensus tree in our simulations) using the HybridSim program of Woodhams et al. (2016). Second, still using HybridSim, we generated $K$-1 gene phylogenies, $T_1, \ldots, T_K$, defined on the same set of $n$ leaves (i.e. they played the role of the other consensus trees in our simulations). Each of these phylogenies differed from $T_0$ by a specific number of hybridization events (the value of the *hybridization_rate* parameter in HybridSim varied from 1 to 4 in our experiments; this value was drawn randomly using a uniform distribution). The number of clusters, $K$, in this experiment varied from 1 to 5, and the number of tree leaves, $n$, was taking the values 8, 16, 32, 64 (and 128, when the comparison with the state-of-the-art tree partitioning algorithms was carried out). The HybridSim program allows one to generate phylogenies in the presence of hybridization and coalescence/incomplete lineage sorting events. Thus, we used the *hybridization_rate* parameter of HybridSim to generate centers of gene tree clusters (i.e. multiple consensus trees or multiple supertrees) and the *coalescence_rate* parameter to generate incongruence across gene trees. The value of the *coalescence_rate* parameter, adding some noise to gene phylogenies, varied between 1 (high noise) and 10 (low noise); the other HybridSim parameters were the default program parameters. Third, for each gene phylogeny $T_i$ ($i = 1, ..., K$), being the center of cluster $i$, we randomly generated a set of 100 trees $T_i'$ (the number of trees $T_i'$ varied form 20 to 100 with the step of 20, and an extra experiment was conducted with datasets with 250 trees, in the simulation conducted with the state-of-the-art tree partitioning methods; see Fig. 2 in the main text), belonging to cluster $i$, such that each tree $T_i'$ differed from $T_i$ by a specific number of coalescence/incomplete lineage sorting patterns of incongruence, which was controlled through the value of the *coalescence_rate* parameter.

First, we compared the classification performances (in terms of Adjusted Rand Index, ARI), of the six variants of our tree clustering algorithm applied to trees with identical sets of leaves (see Fig. 1). The six evaluated variants of our algorithm were based on: (1) the $OF_{EA}$ objective function with approximation by Euclidean distance and *CH* cluster validity index (*CH-E*), (2) the $OF_{LA}$ objective function with approximation by the lower bound and *CH* cluster validity index (*CH-LB*), (3) the $OF_{MA}$ objective function with approximation by the mean of interval and *CH* cluster validity index (*CH-MI*), (4) the $OF_{EA}$ objective function and Silhouette cluster validity index (*SH*), (5) the $OF_{EA}$ objective function and the Gap cluster validity index (*Gap*), and (6) $OF_{EA}$ objective function and Ball and Hall cluster validity index (*BH*).

Second, the variant of our algorithm based on the $OF_{EA}$ objective function with approximation by Euclidean distance and *CH* cluster validity index (*CH-E*) that showed the best overall per-formance in the first simulation was compared to the state-of-the-art tree clustering methods, including: (1) the MPC method by Bonnard et al. (2006), (2) the tree clustering algorithm by Stockham et al. (2002), which is based on $k$-means clustering with squared $RF$ distance (this method recomputes the majority-rule consensus trees of all clusters at each $k$-means iteration), and (3) the $k$-medoids tree clustering algorithm by Tahiri et al. (2018) , which uses the $RF$ distance and $SH$ cluster validity index. The comparison was conducted in terms of the quality of clustering results returned by competing methods (Fig. 2a and b) and the running time (Fig. 2c and d).

Our second simulation experiment involved partitioning trees with different sets of leaves with the objective to build multiple consensus supertrees. The data generation protocol for this experiment included an additional step consisting of the random removal of some species (i.e. tree leaves) from the generated gene trees. The branches adjacent to the removed leaves were collapsed. The following intervals of missing data were considered: 0% (no species were removed), 10% (5% to 15% of species were randomly removed), 25% (16% to 35% of species were randomly removed) and 50% (36% to 65% of species were randomly removed). The exact number of species to be removed from each gene tree and each data interval was drawn randomly using a uniform distribution. We also made sure that every pair of trees in each input dataset had at least 4 species in common. The value of the penalization parameter $\alpha$ in Equation 10 was set to 0 in this experiment. Two independent simulations were carried for the supertree version of our algorithm using the $OF_{ST\_EA}$ objective function (Equation 10) and the $CH$-$E$ (Equations 15-20) and $BH$ cluster validity indices adapted for supertree partitioning (see Supplementary Figs. S3 and S4). The $OF_{ST\_EA}$ objective function was used because it provided the best overall performance in our first simulation experiment with consensus trees (see Fig. 1), whereas the $BH$ index was used because it slightly outperformed the $Gap$ index in case of heterogeneous data (i.e. when the number of clusters $K$ was equal to 1).

Our third simulation experiment was also conducted to evaluate the ability of our algorithm to cope with incomplete data. As in the second experiment, gene trees with different sets of leaves were considered. The supertree version of our algorithm based on the $OF_{ST\_EA}$ objective function (Equation 10) and the $CH$-$E$ cluster validity index (Equations 15-20) was used here with different values of the penalization parameter $\alpha$, which varied from 0 to 1 (with the step of 0.2; see Fig. 3).

In all simulation experiments, our tree partitioning algorithm was carried out with 100 random starts until the convergence of the selected objective function or until 50 iterations in the algorithm's internal loop were completed (i.e. the same stopping rule as in the traditional *k*-means algorithm were applied). All reported ARI results (see Figs 1 to 3 and Supplementary Figs. S3-S4) are the averages taken over all considered numbers of trees, leaves and clusters. The simulation results presented in Figures 1 and Supplementary Figs. S3 and S4 (portions a, b and d, in all these figures) and Figures 2 and 3 correspond to the case where the value of the coalescence rate parameter was fixed to 5. Figure 1 and Supplementary Figs. S3 and S4 (portion c, in all these figures) illustrate how the algorithm's results vary with respect to the change in the coalescence rate.

Our computational experiments were carried out using a 64-bit PC computer equipped with an Intel i7-8750H CPU (2.5 GHz) and 32 Gb of RAM, except for the simulation comparing the performances of the state-of-the-art clustering algorithms, which was conducted on a high-performance parallel computing server of Compute Canada.

**Appendix S5. Simulations with multiple alternative supertrees using *CH* and *BH* indices.**

Supplementary Figures S3 and S4 illustrate the classification performance of our supertree clustering algorithm based on the $OF_{ST\_EA}$ objective function used with *CH* and *BH* cluster validity indices, respectively. Here, the value of the penalization parameter $\alpha$ in Equation 10 was set to 0.

The algorithm was applied to gene trees containing different numbers and sets of leaves. We can observe that the clustering performances of both tested algorithm's variants gradually decreases as the number of missing species (i.e. tree leaves) increases. The supertree clustering algorithm based on *CH* generally outperformed that based on *BH*, but the *BH*-based variant seems to be less sensitive to the increase in the number of missing species as the number of clusters and the coalescence noise grow (e.g. for the case of 5 tree clusters, the average ARI value provided by the *CH*-based version decreased from 0.97 for 0% of missing species to 0.81 for 50% of missing species (Fig. S3a), whereas for the *BH*-based version it decreased from 0.51 for 0% of missing species to 0.44 for 50% of missing species (Fig. S4a)).

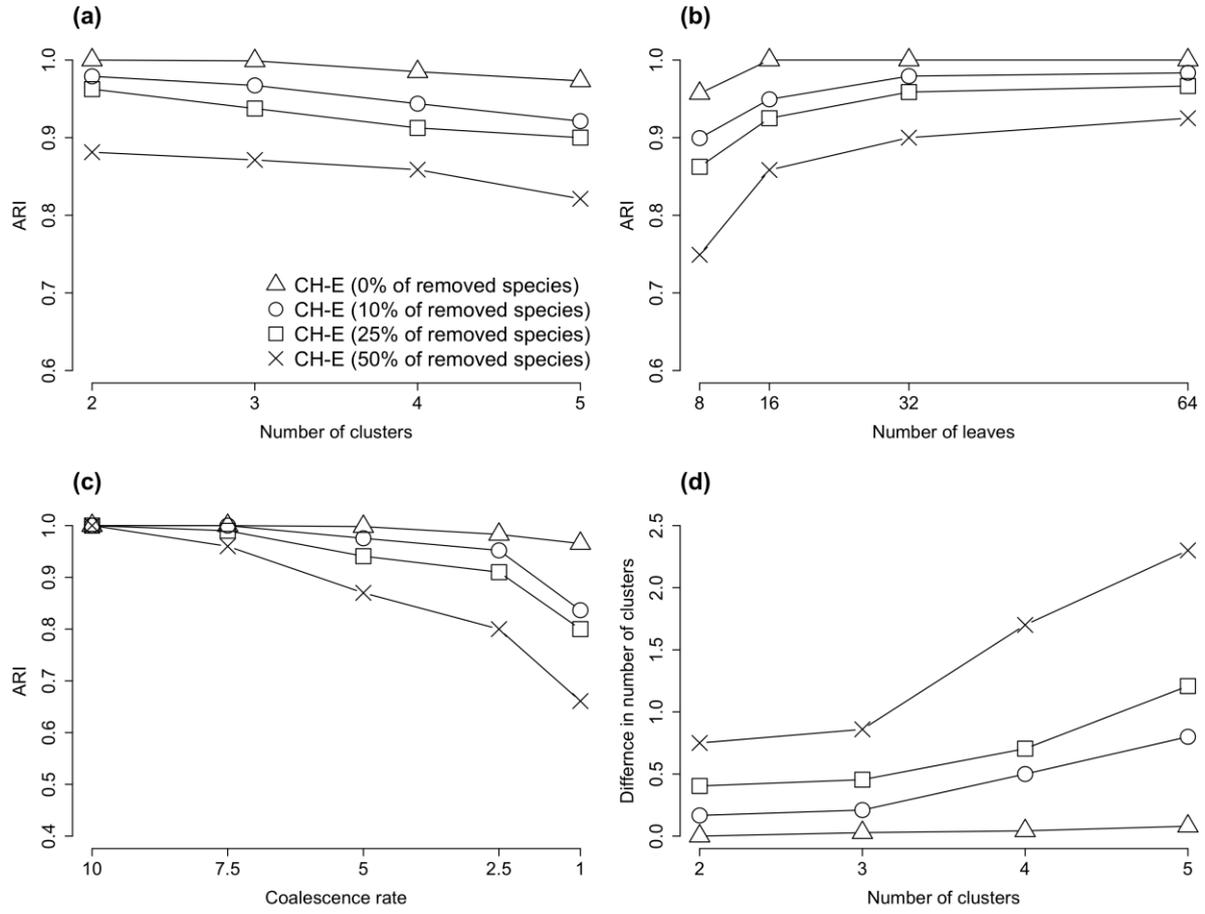

**Supplementary Fig. S3.** Classification performance of our *k*-means-based supertree clustering algorithm applied to trees with different numbers and sets of leaves (the following numbers of leaves were randomly removed from each tree: 0% (no species were removed), 10% (5% to 15% of species were removed), 25% (16% to 35% of species were removed) and 50% (36% to 65% of species were removed)) using the $OF_{ST\_EA}$ objective function with approximation by Euclidean distance and the *CH* cluster validity index (*CH-E*). The value of the penalization parameter $\alpha$ was set to 0. The results are presented in terms of *ARI* with respect to the: (a) number of tree clusters, (b) number of leaves and (c) coalescence rate; (d) average absolute difference between the true and the obtained number of clusters. The coalescence rate parameter in the HybridSim program was set to 5 in simulations (a), (b) and (d). The presented results are the averages taken over all considered numbers of leaves in simulations (a), (c) and (d), and all considered numbers of clusters in simulations (b) and (c).

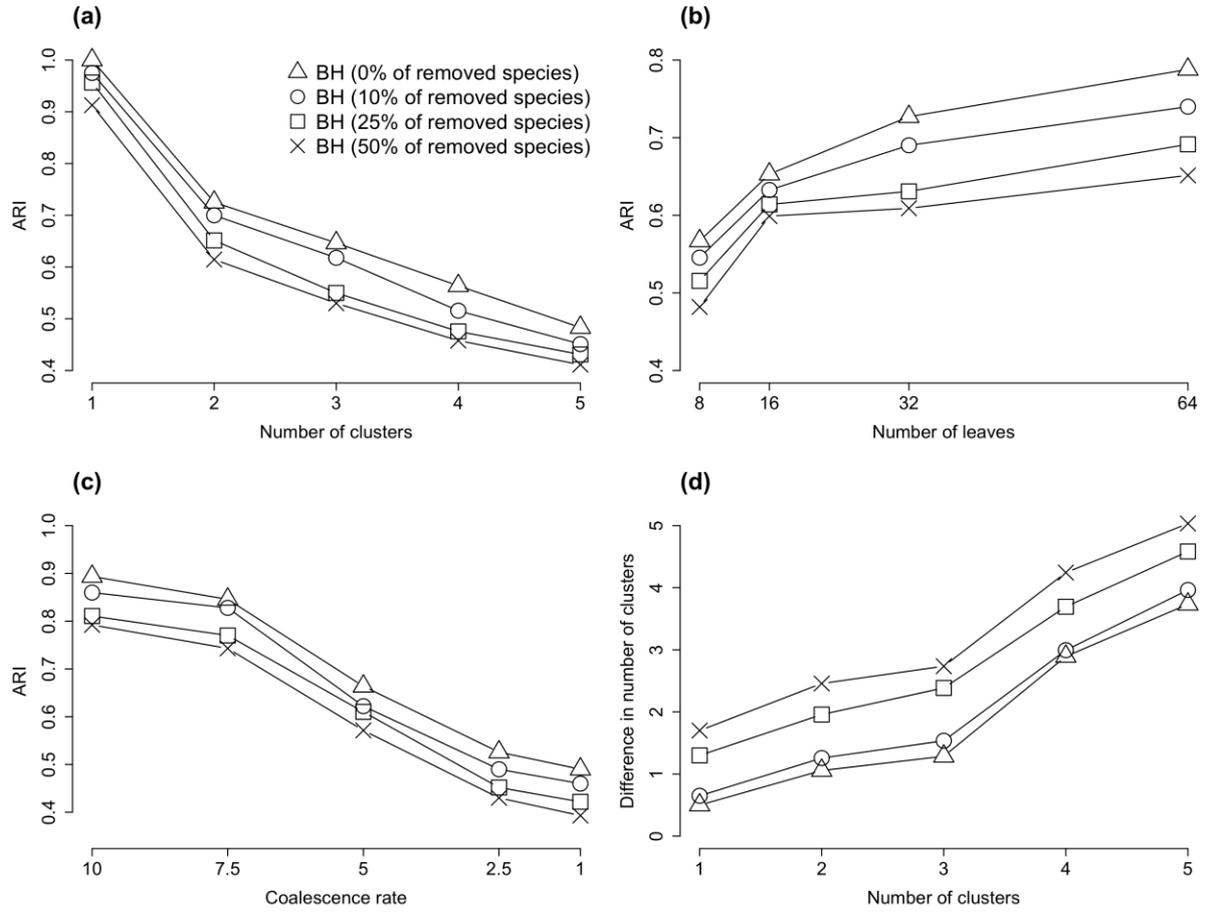

**Supplementary Fig. S4.** Classification performance of our *k*-means-based supertree clustering algorithm applied to trees with different numbers and sets of leaves using the $OF_{ST\_EA}$ objective function with approximation by Euclidean distance and the Ball and Hall cluster validity index (*BH*). **Supplementary** Figure S3 panel description applies here.

**Appendix S6. Exploring the patterns of evolution of SARS-CoV-2 genes.**

In this Appendix, we first describe the SARS-CoV-2 dataset examined in our work, then give some details regarding the applied multiple sequence alignment and tree inference methods, and finally present and discuss the results of our supertree clustering analysis (see Fig. 4 and the related description in the main text for the presentation of the obtained results). Several recent studies provide evidence of recombination events in different genes of betacoronavirus organisms. For example, Boni et al. (2020) pointed out that sarbecoviruses (i.e. the viral sub-genus containing SARS-CoV and SARS-CoV-2) undergo frequent recombination events and exhibit spatially structured genetic diversity on a regional scale in China. Li et al. (2020) demonstrated that SARS-CoV-2's receptor binding domain has been introduced through re-combination with a pangolin coronavirus and indicated that similar purifying selection in different host species, along with frequent recombination among coronaviruses, may represent a common evolutionary mechanism leading to the emergence of human coronaviruses.

*Data description*

To carry out our the supertree clustering analysis, we considered the case of evolution of 43 betacoronavirus organisms, including : (1) four SARS-CoV-2 genomes from China, Australia, Italy and USA, coming from different clades of the Gisaid SARS-CoV-2 phylogeny (see https://www.gisaid.org; Shu and McCauley 2017); (2) the RaTG13 bat CoV genome from *Rhinolophus affinis* (data collected in the Yunnan province of China); (3) five Guangxi (GX) pangolin CoV genomes (data provided by the Beijing Institute of Microbiology and Epidemiology); (4) two Guangdong (GD) pangolin CoV genomes (data extracted from dead Malayan pangolins in the Guangdong province of China); (5) the bat CoV ZC45 and ZXC21 genomes (i.e. bat CoVZ clade with data coming from the Zhejiang province of China); (6) five additional CoV genomes extracted from bats across different provinces of China (i.e. BtCoV 273 2005, Rf1, HKU3-12, HKU3-6 and BtCoV 279 2005 CoVs); (7) four SARS-CoV strains related to the first SARS outbreak (i.e. SARS, Tor2, SARS-CoV BJ182-4 and bat Rs3367 CoV found in *Rhinolophus sinicus*); (8) the BtKY72 and BM48 31 BGR 2008 CoV genomes, extracted from bats in Kenya and Bulgaria; (9) the MERS-CoV and the related bat HKU-4 and HKU-5 CoV genomes; (10) a human HKU1 CoV genome; (11) a feline CoV genome; (12) four murine CoV and the related Rat Parker CoV genomes; (13) an equine CoV genome; (14) a porcine CoV genome; (15) the rabbit HKU14 CoV genome; (16) human enteric and human

OC43 CoV genomes; (17) three bovine CoV genomes, including AH187 and OH440 bovine CoVs. The first 25 of these CoV genomes (sub-groups 1 to 8 above) include the closest relatives of SARS-CoV-2; they have been originally examined in Lam et al. (2020). The remaining 18 CoVs (sub-groups 9 to 17 above) comprise betacoronaviruses labeled as common cold CoVs in the Gisaid coronavirus tree (Shu and McCauley 2017) and those studied by Prabakaran et al. (2006). Supplementary Table S1 (see Supplementary Material) provides the organism names, host species, and Gisaid or GenBank accession numbers for the 43 coronaviruses considered in our study.

### Methods details

Our analysis was conducted for 11 main genes of the SARS-Cov-2 genome (i.e. genes ORF1ab, S, ORF3a, E, M, ORF6, ORF7a, ORF7b, ORF8, N and ORF10) as well as for the RD domain of the spike protein because of its key evolutionary importance. Thus, 12 gene phylogenies were inferred and analyzed in our study. It is worth mentioning that some gene annotations were absent in the GenBank and Gisaid databases (i.e. annotations for genes ORF6, ORF7a, ORF7b, ORF8 and ORF10 for taxa from sub-groups 9 to 17 and annotations for gene ORF8 for taxa from sub-group 8 were unavailable), thus leading to gene phylogenies having different, but mutually overlapping, sets of leaves.

The VGAS program (Zhang et al. 2019), intended to identify viral genes and carry out gene function annotation, was executed to validate all betacoronavirus genes found in GenBank and Gisaid. We then performed multiple sequence alignments (MSAs) for the 11 considered coronavirus genes (DNA sequences) and for the RB domain (amino acid sequences) by means of the MUSCLE algorithm (Edgar 2004) using the default parameters of the MegaX program (v. 10.1.7) (Kumar et al. 2018). The obtained MSAs were used to build gene trees presented in Supplementary Figures S5 to S16 (see Supplementary Material). Moreover, in order to apply the HGT and recombination detection methods on the obtained gene and consensus trees, we inferred a species phylogeny (see Figure 4d in the main text) of the 43 considered CoVs using the same version of MUSCLE. The GBlocks algorithm (v. 0.91b, Castresana 2000), available at the Phylogeny.fr web server (Dereeper et al. 2008), was then used with the less stringent correction option to remove MSA sites with large gap ratios. Gene and genome trees that will be further used in clustering and HGT (Horizontal Gene Transfer) and recombination analyses were inferred using the RAxML algorithm (v. 0.9.0; Stamatakis 2006). The most suitable DNA/amino acid substitution model determined by MegaX, and available on the RAxML web site (https://raxml-ng.vital-it.ch), was used for each MSA. Precisely, the (GTR+G+I) model was found to be the best-fit substitution model for genes ORF1ab, S, N and for the whole genomes, the (HKY+G) model was the most suitable for genes ORF3a, E, ORF6, ORF7a, the (HKY+I) model for gene ORF7b, the (HKY+G+I) for gene ORF8, the (JC) model for gene ORF10, and the (WAG+G) model for the RB domain. The tree inference was conducted using the bootstrap option (with 100 replicates for each MSA considered).

The consensus supertrees (see Supplementary Figures S4a, b and c or Figures 4a, b and c in the main text) for each cluster found by our algorithm were inferred using the CLANN program (Creeve and McInerney 2005). The HGT-Detection program (Boc et al. 2010) from the Armadillo 1.1 workflow platform (Lord et al. 2012) was used to infer directional horizontal gene transfer-recombination network (Huson and Bryant 2005; Beiko et al. 2005; Boc et al. 2012) for the three obtained consensus supertrees (see Figure 4d in the main text).

Multiple sequence alignments for all gene and genome sequences used in this study as well as all inferred gene and species phylogenies (in the Newick format) are available at: http://www.info2.uqam.ca/~makarenkov_v/Supplementary_Material_data.zip.

**Supplementary Table S1. Full virus names, abbreviations, host species and GenBank/GISAID accession numbers for the 43 betacoronavirus genomes analysed in our study.**

| Organism's name | Abbreviation | Host | Accession number (Gen-Bank, GISAID) |
|---|---|---|---|
| hCoV-19/Australia/VIC231/2020 | Hu Australia VIC231 2020 | Human | EPI_ISL_419 926 |
| hCoV-19/USA/UT-00346/2020 | Hu USA UT 00346 20202 | Human | EPI_ISL_420 819 |
| BetaCoV/Wuhan-Hu-1 | Hu Wuhan 2020 | Human | NC_045512.2 |
| hCoV-19/Italy/TE4836/2020 | Hu Italy TE4836 2020 | Human | EPI_ISL_418 260 |
| Bat coronavirus RaTG13 | RaTG13 | Bat | MN996532.1 |
| hCoV-19/pangolin/Guangdong/1/2019 | Guangdong Pangolin 1 2019 | Pangolin | EPI_ISL_410 721 |
| hCoV-19/pangolin/Guangdong/P2S/2019 | Guangdong Pangolin P2S 2019 | Pangolin | EPI_ISL_410 544 |
| PCoV_GX-P5E | Guangxi Pangolin P5E | Pangolin | MT040336 |
| PCoV_GX-P2V | Guangxi Pangolin P2V | Pangolin | MT072864 |
| PCoV_GX-P5L | Guangxi Pangolin P5L | Pangolin | MT040335 |
| PCoV_GX-P1E | Guangxi Pangolin P1E | Pangolin | MT040334.1 |
| PCoV_GX-P4L | Guangxi Pangolin P4L | Pangolin | MT040333 |
| bat-SL-CoVZC45 | Bat CoVZC45 | Bat | MG772933.1 |
| bat-SL-CoVZXC21 | Bat CoVZXC21 | Bat | MG772934.1 |
| BtCoV/273/2005 | BtCoV 273 2005 | Bat | DQ648856.1 |
| Bat SARS coronavirus Rf1 | Rf1 | Bat | DQ412042.1 |
| Bat SARS coronavirus HKU3-12 | HKU3-12 | Bat | GQ153547.1 |
| Bat SARS coronavirus HKU3-6 | HKU3-6 | Bat | GQ153541.1 |
| BtCoV/279/2005 | BtCoV 279 2005 | Bat | DQ648857.1 |
| SARS coronavirus BJ01 | SARS | Human | AY278488.2 |
| SARS coronavirus | Tor2 | Human | NC_004718.3 |

| SARS coronavirus BJ182-4 | SARS-CoV BJ182-4 | Human | EU371562 |
|---|---|---|---|
| Bat SARS-like coronavirus Rs3367 | Rs3367 | Bat | KC881006.1 |
| SARS-related coronavirus BtKY72 | BtKY72 | Bat | KY352407.1 |
| Bat coronavirus BM48-31/BGR/2008 | BM48 31 BGR 2008 | Bat | GU190215.1 |
| Human betacoronavirus 2c EMC/2012 | MERS-CoV S | Human | JX869059 |
| Bat coronavirus HKU5-1 | Bat HKU5-1 | Bat | NC_009020 |
| Bat coronavirus HKU4-1 | Bat HKU4-1 | Bat | NC_009019 |
| Feline infectious peritonitis virus | Feline per | Cat | NC_002306 |
| Human coronavirus HKU1 | HKU1 | Human | NC_006577 |
| Murine coronavirus RA59/R13 | Murine RA59/R13 | Mouse | ACN89689 |
| Murine hepatitis virus strain 4 | Murine hep 4 | Mouse | P22432 |
| Mouse hepatitis virus strain MHV-A59 C12 mutant | Murine A59 | Mouse | NC_001846 |
| Murine hepatitis virus | Murine virus | Mouse | ABS87264 |
| Rat coronavirus Parker | Rat Parker | Rat | NC_012936 |
| Rabbit coronavirus HKU14 | Rabbit HKU14 | Rabbit | NC_017083 |
| Equine coronavirus | Equine NC99 | Horse | NC_010327 |
| Porcine hemagglutinating en-cephalomyelitis virus | Porcine v | Pig | NC_007732 |
| Human coronavirus OC43 | Human OC43 | Human | NC_005147 |
| Human enteric coronavirus strain 4408 | Human ent 4408 | Human | NC_012950 |
| Bovine coronavirus | Bovine CoV | Calf | NC_003045 |
| Bovine respiratory coronavirus AH187 | Bovine AH187 | Calf | NC_012948 |
| Bovine respiratory coronavirus bovine/US/OH-440-TC/1996 | Bovine OH440 | Calf | NC_012949 |

**Tree for gene ORF1ab (43 taxa)**

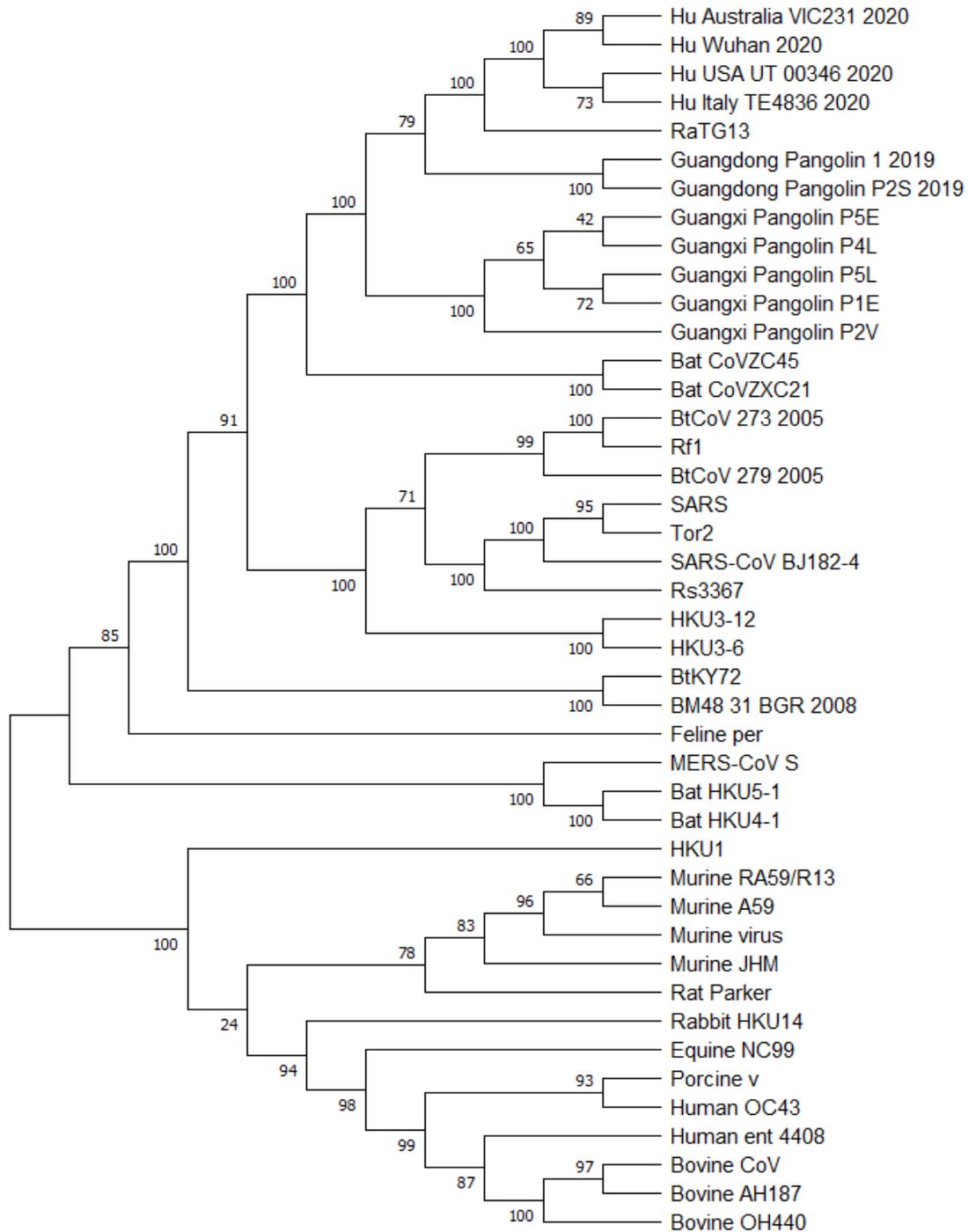

**Supplementary Fig. S5. Phylogenetic tree of gene ORF1ab inferred using RAxML with 100 bootstrap replicates for a group of 43 betacoronaviruses (see S1 Table).**

**Tree for gene S (43 taxa)**

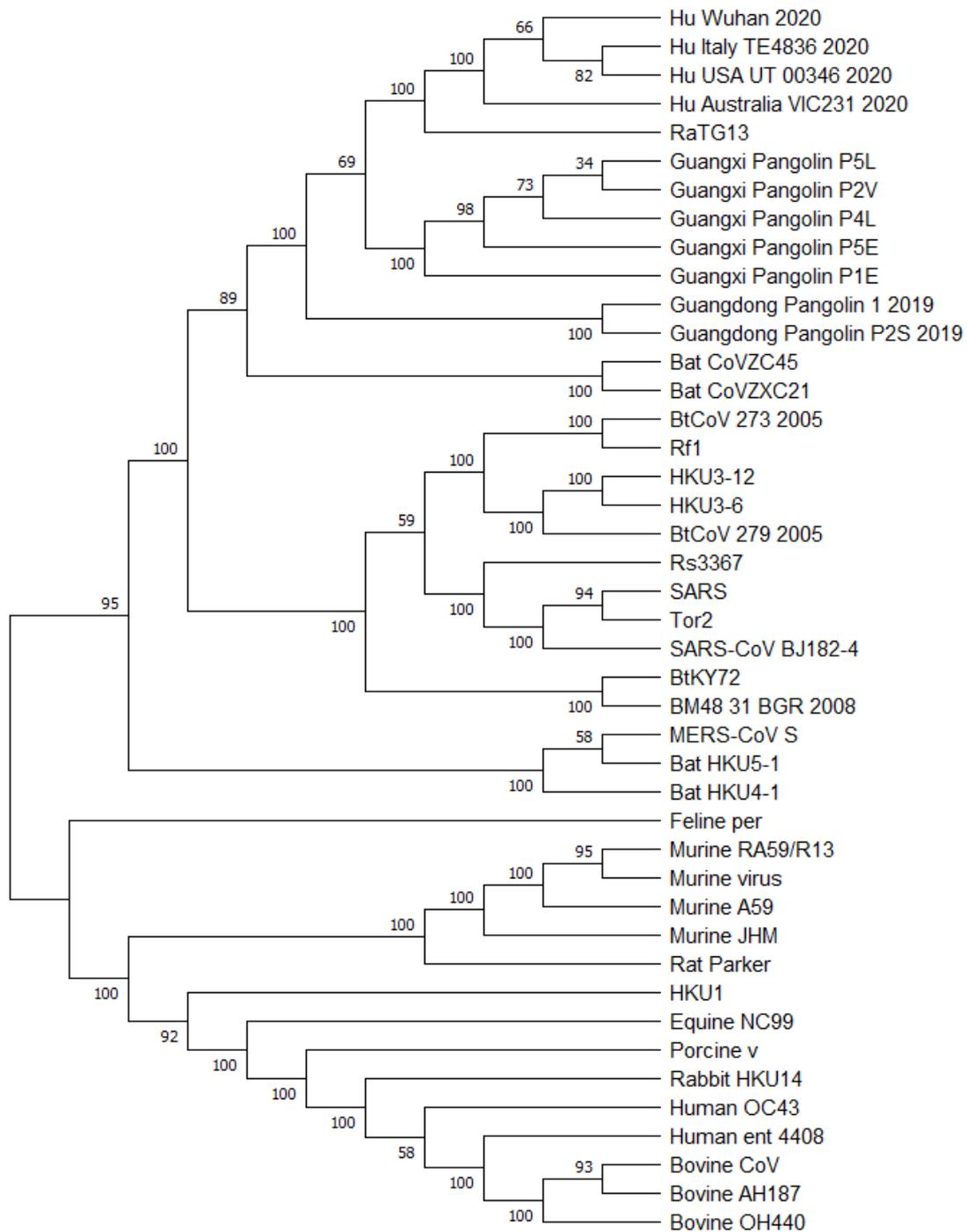

**Supplementary Fig. S6. Phylogenetic tree of gene S inferred using RAxML with 100 bootstrap replicates for a group of 43 betacoronaviruses (see S1 Table).**

**Tree for gene ORF3a (43 taxa)**

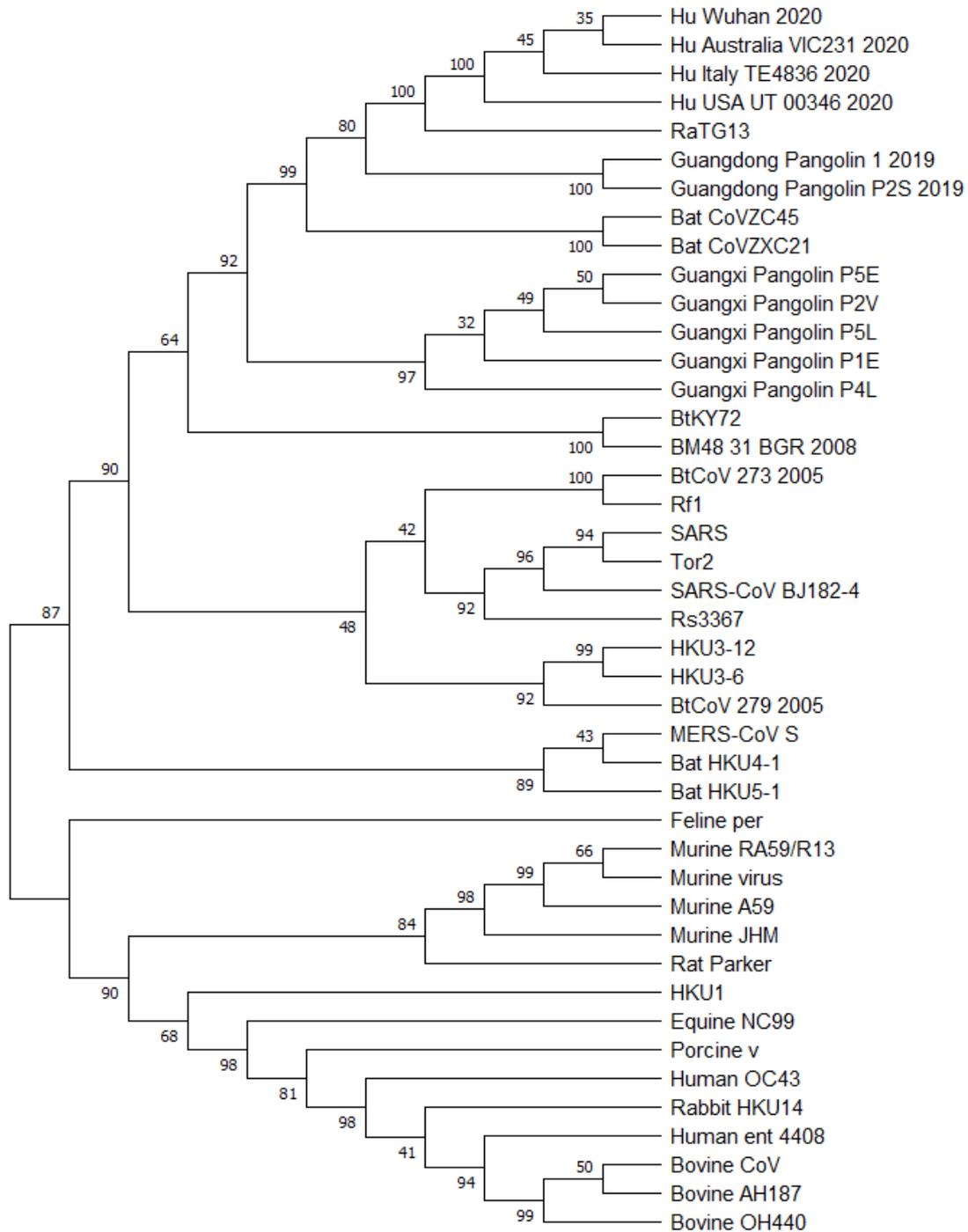

**Supplementary Fig. S7. Phylogenetic tree of gene ORF3a inferred using RAxML with 100 bootstrap replicates for a group of 43 betacoronaviruses (see S1 Table).**

**Tree for gene E (43 taxa)**

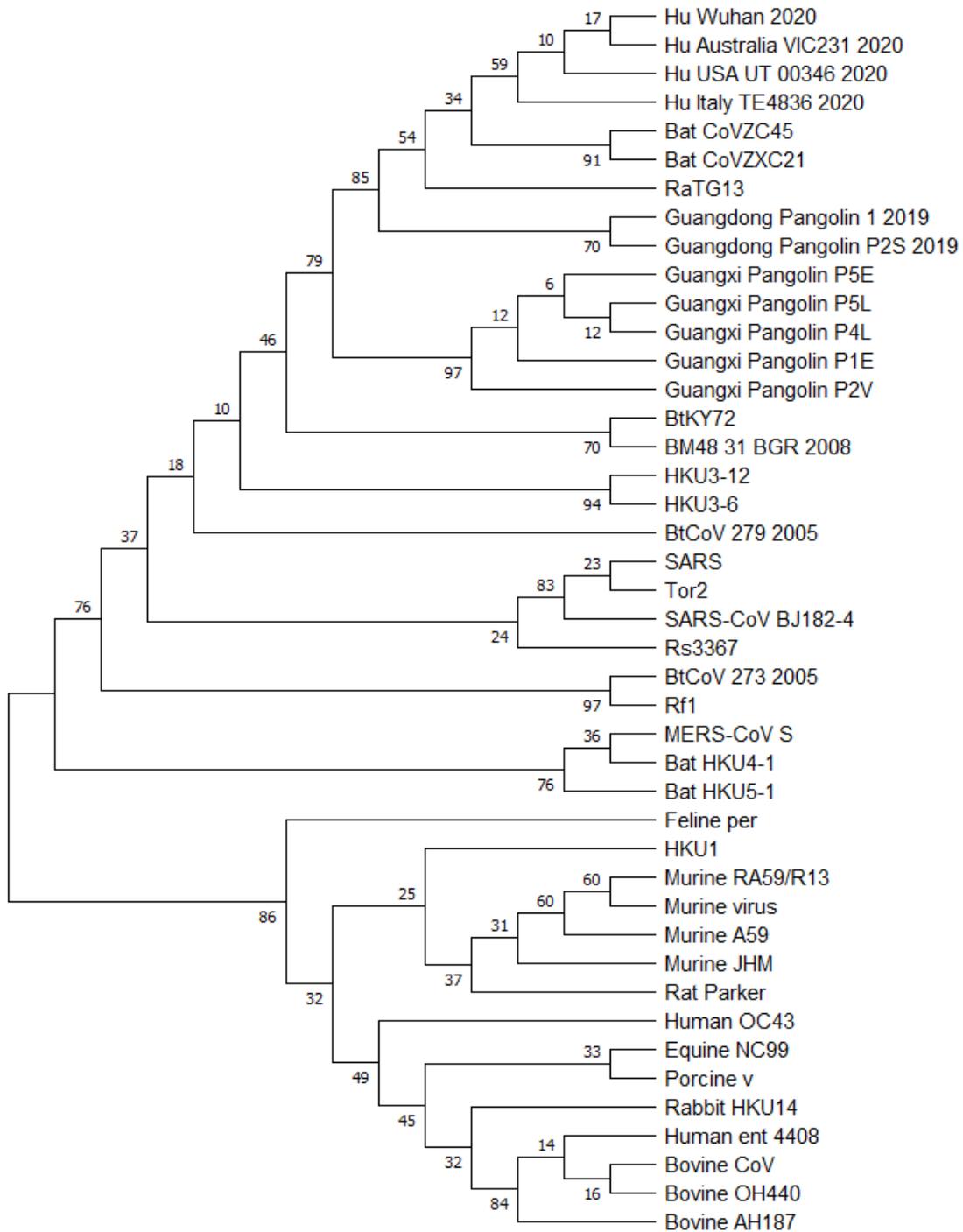

**Supplementary Fig. S8. Phylogenetic tree of gene E inferred using RAxML with 100 bootstrap replicates for a group of 43 betacoronaviruses (see S1 Table).**

**Tree for gene M (43 taxa)**

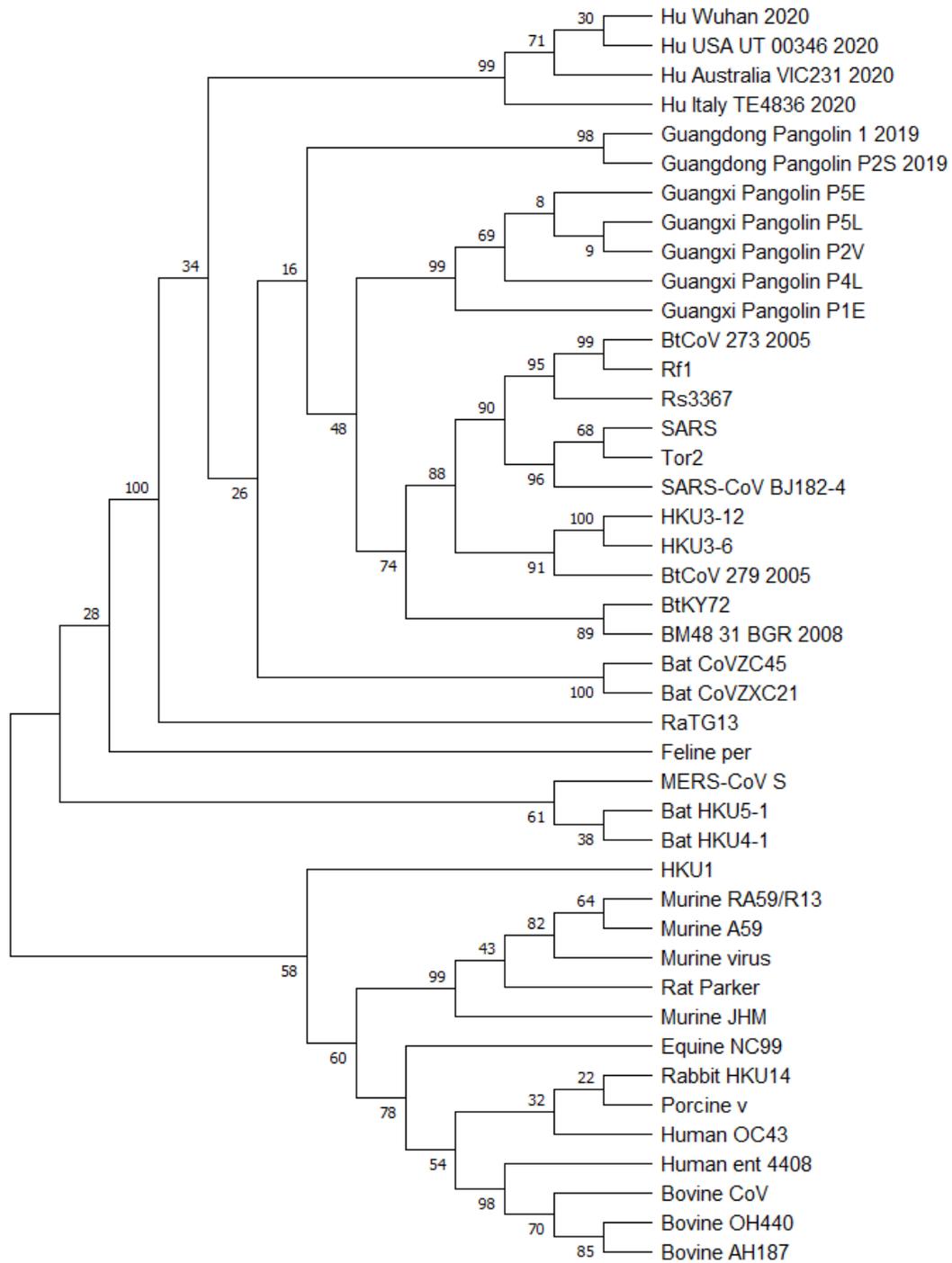

**Supplementary Fig. S9. Phylogenetic tree of gene M inferred using RAxML with 100 bootstrap replicates for a group of 43 betacoronaviruses (see S1 Table).**

**Tree for gene ORF6 (25 taxa)**

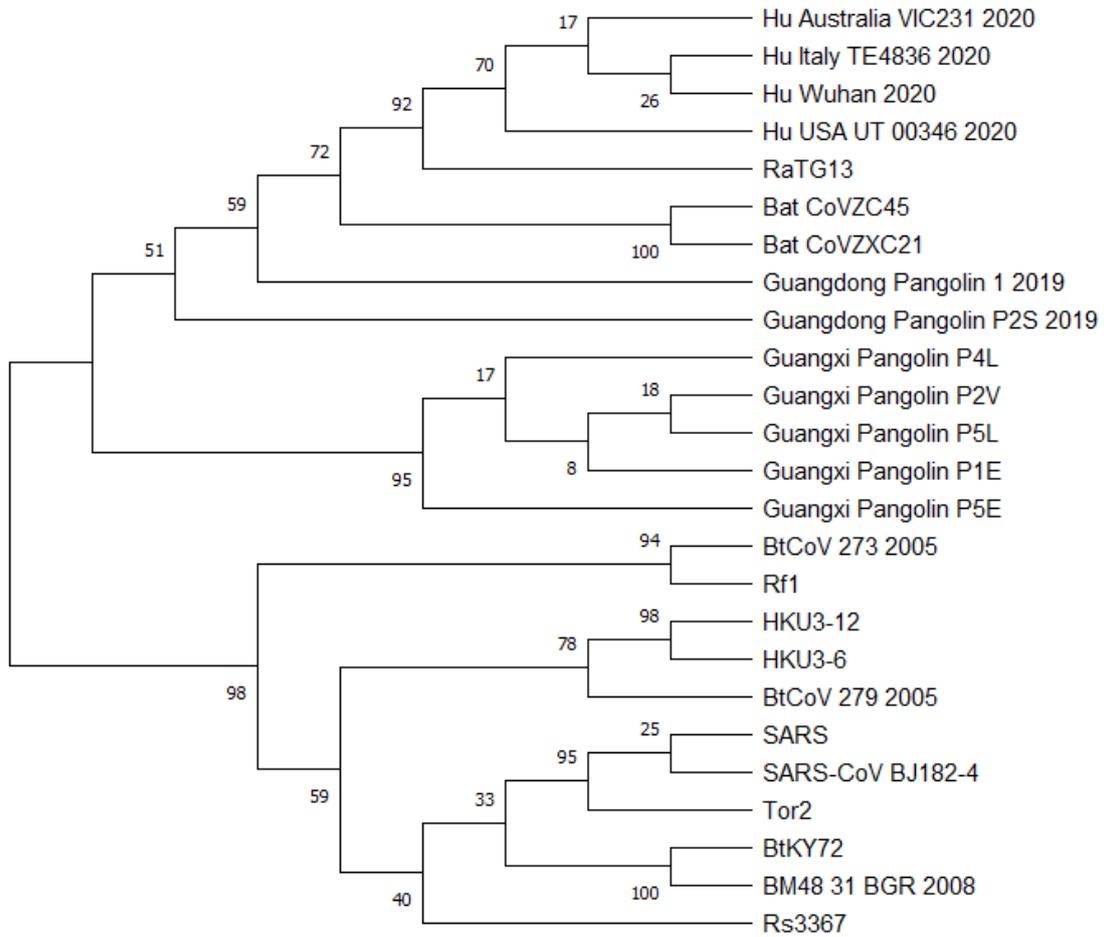

**Supplementary Fig. S10. Phylogenetic tree of gene ORF6 inferred using RAxML with 100 bootstrap replicates for a group of 25 betacoronaviruses (see S1 Table).**

**Tree for gene ORF7a (25 taxa)**

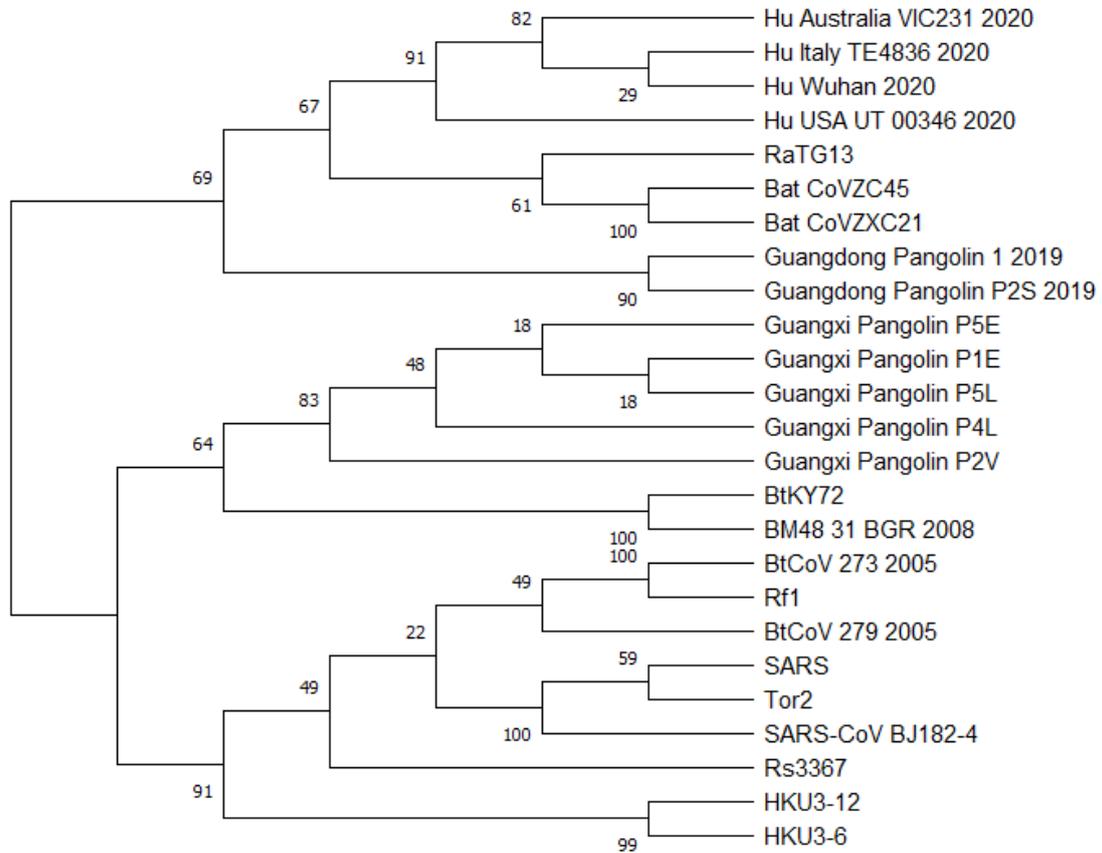

**Supplementary Fig. S11. Phylogenetic tree of gene ORF7a inferred using RAxML with 100 bootstrap replicates for a group of 25 betacoronaviruses (see S1 Table).**

**Tree for gene ORF7b (25 taxa)**

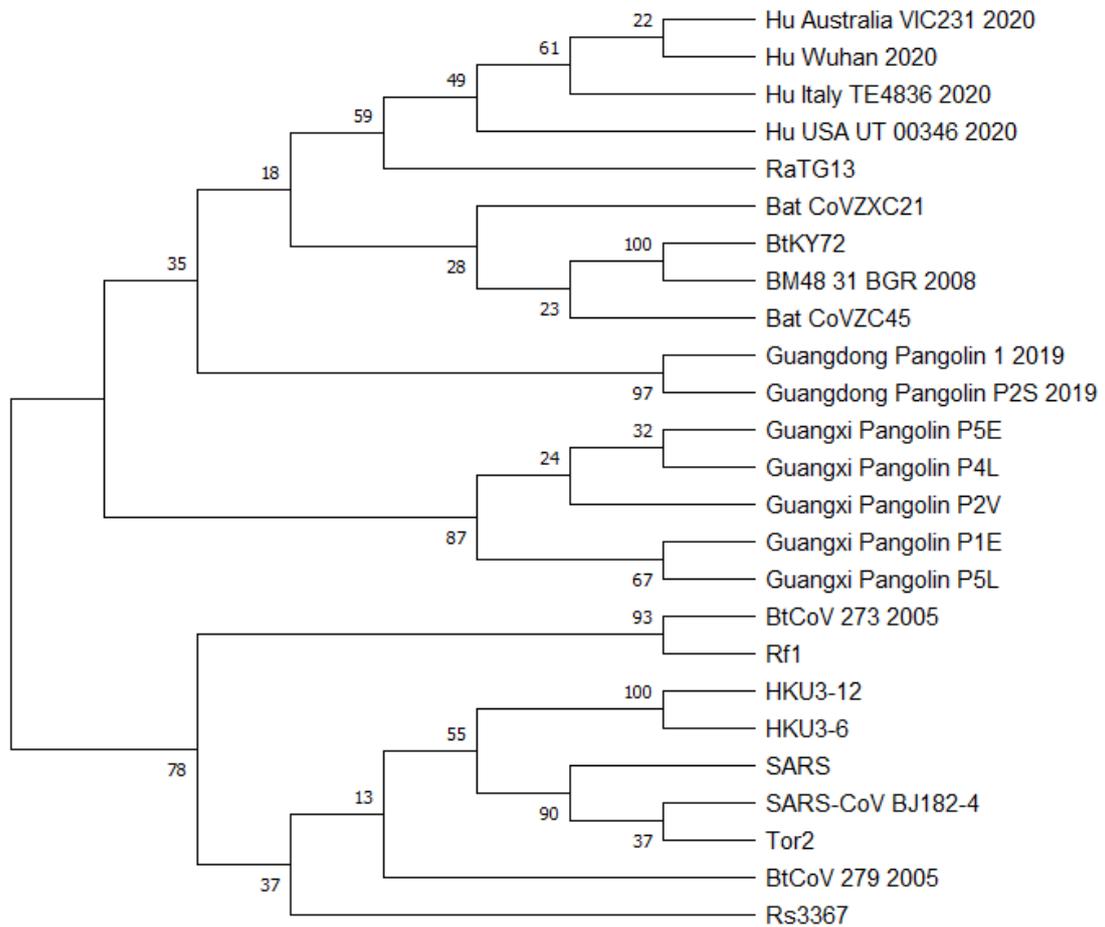

**Supplementary Fig. S12. Phylogenetic tree of gene ORF7b inferred using RAxML with 100 bootstrap replicates for a group of 25 betacoronaviruses (see S1 Table).**

**Tree for gene ORF8 (23 taxa)**

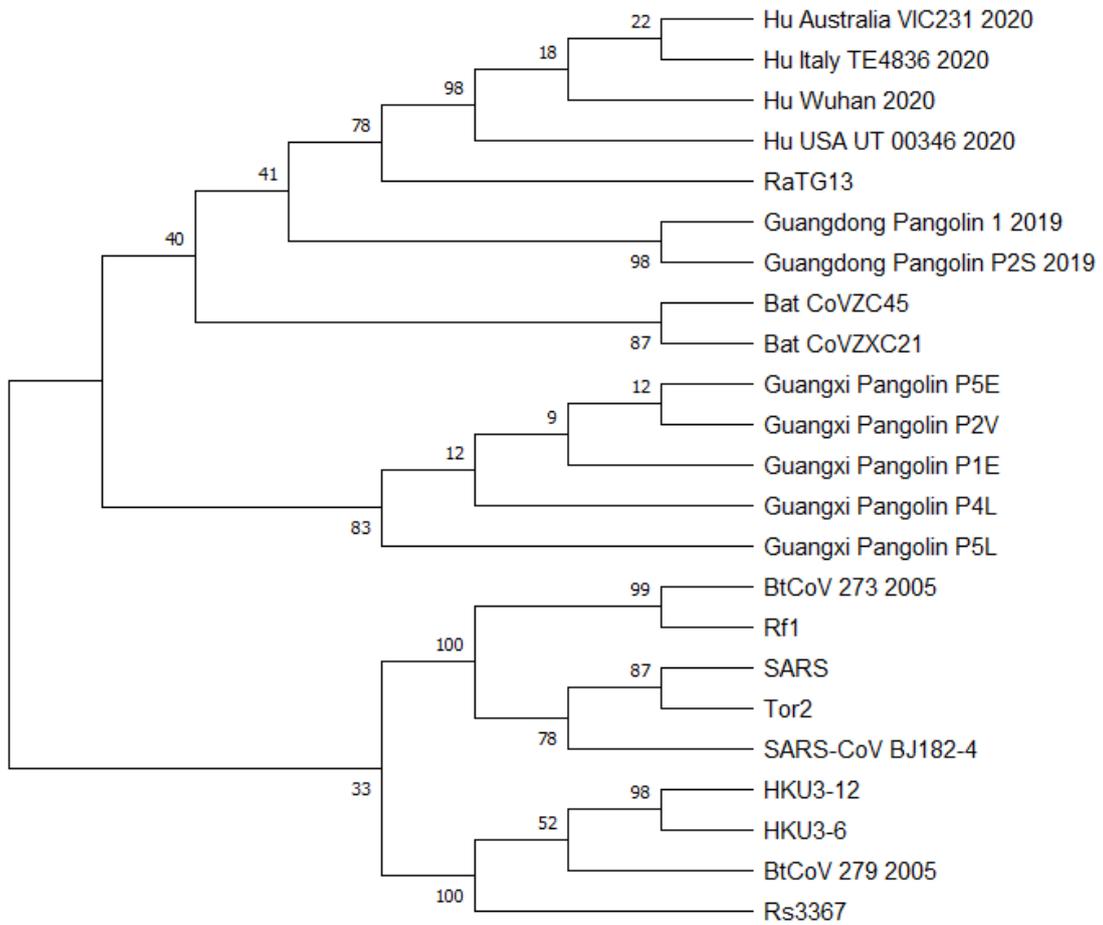

**Supplementary Fig. S13. Phylogenetic tree of gene ORF8 inferred using RAxML with 100 bootstrap replicates for a group of 23 betacoronaviruses (see S1 Table).**

**Tree for gene N (43 taxa)**

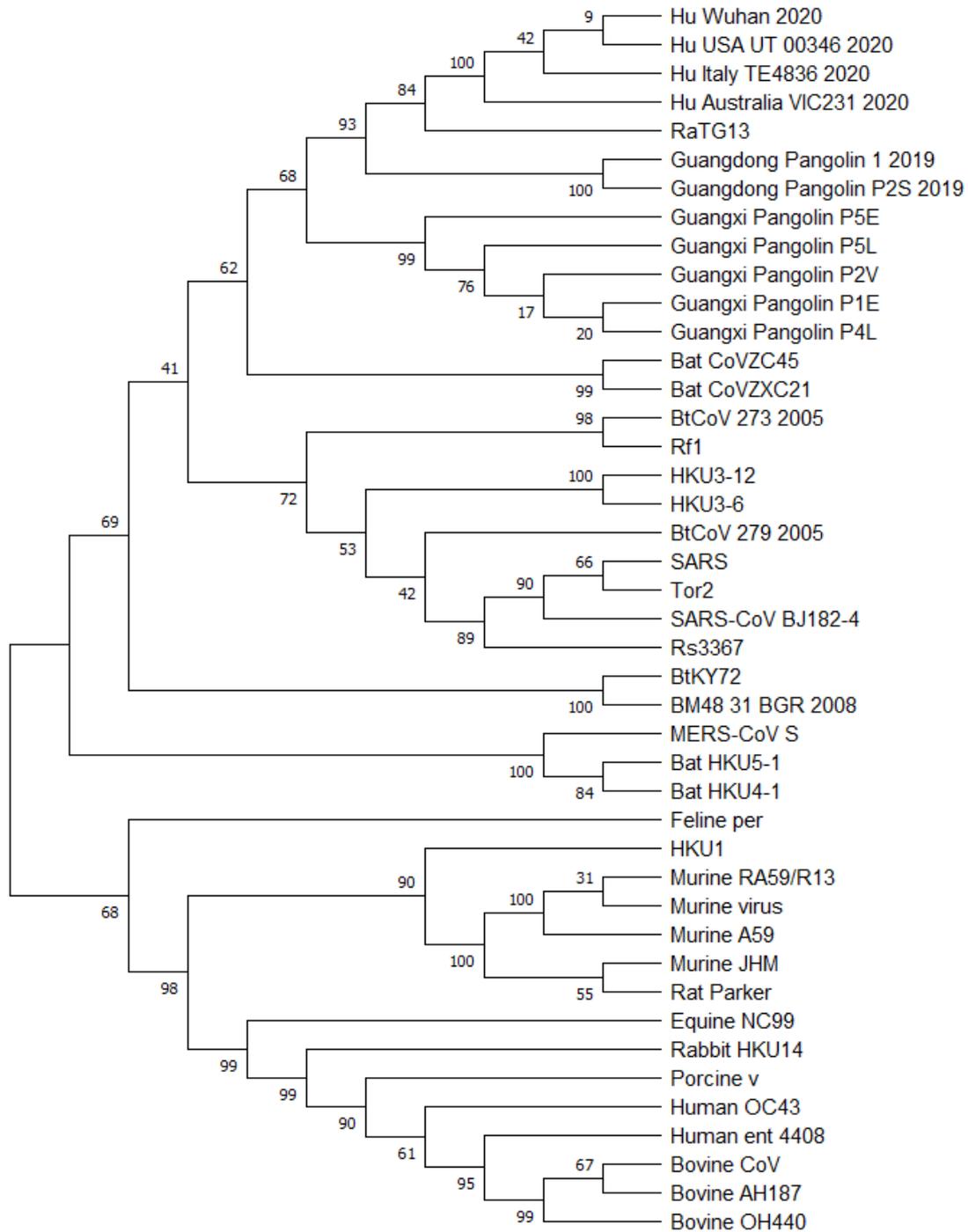

**Supplementary Fig. S14. Phylogenetic tree of gene N inferred using RAxML with 100 bootstrap replicates for a group of 43 betacoronaviruses (see S1 Table).**

**Tree for gene ORF10 (25 taxa)**

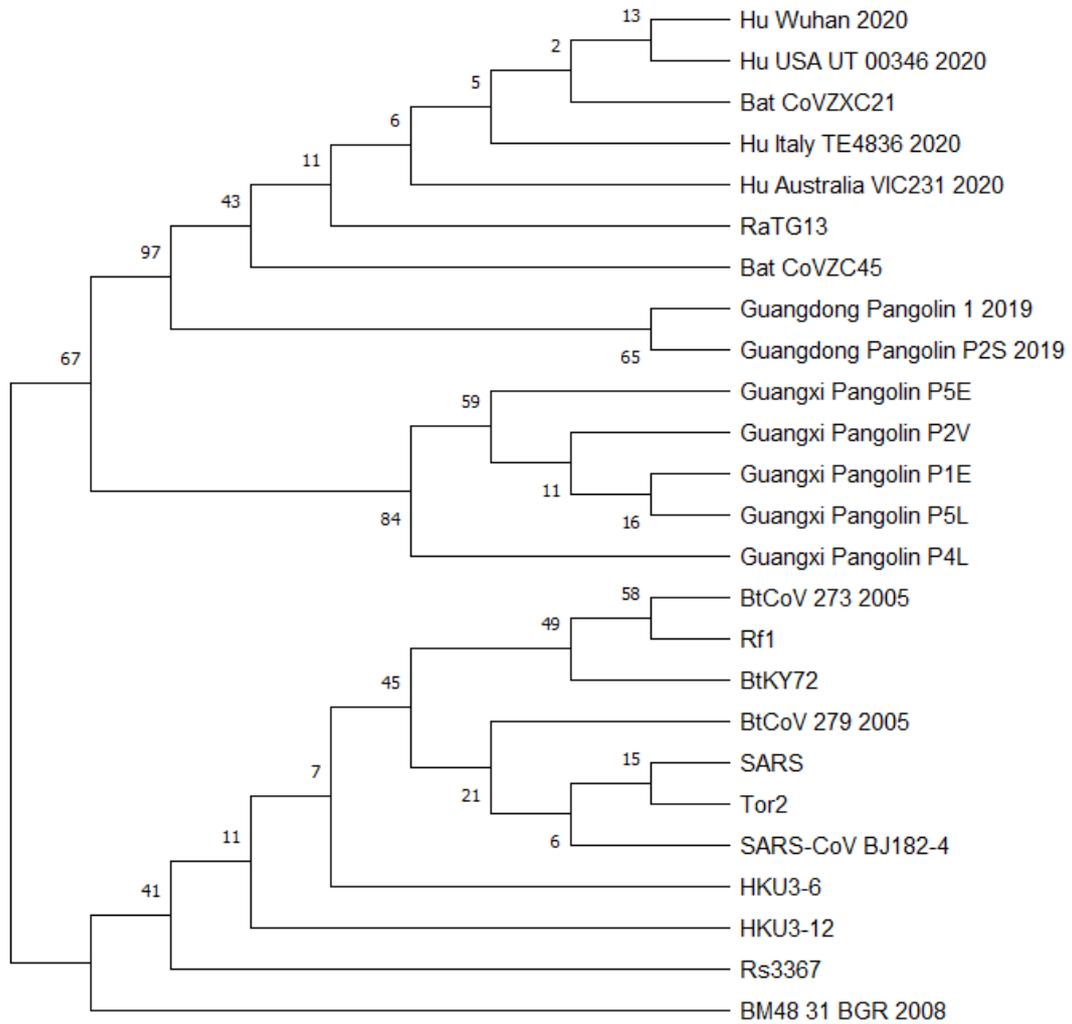

**Supplementary Fig. S15. Phylogenetic tree of gene ORF10 inferred using RAxML with 100 bootstrap replicates for a group of 25 betacoronaviruses (see S1 Table).**

**Tree for RB domain (43 species)**

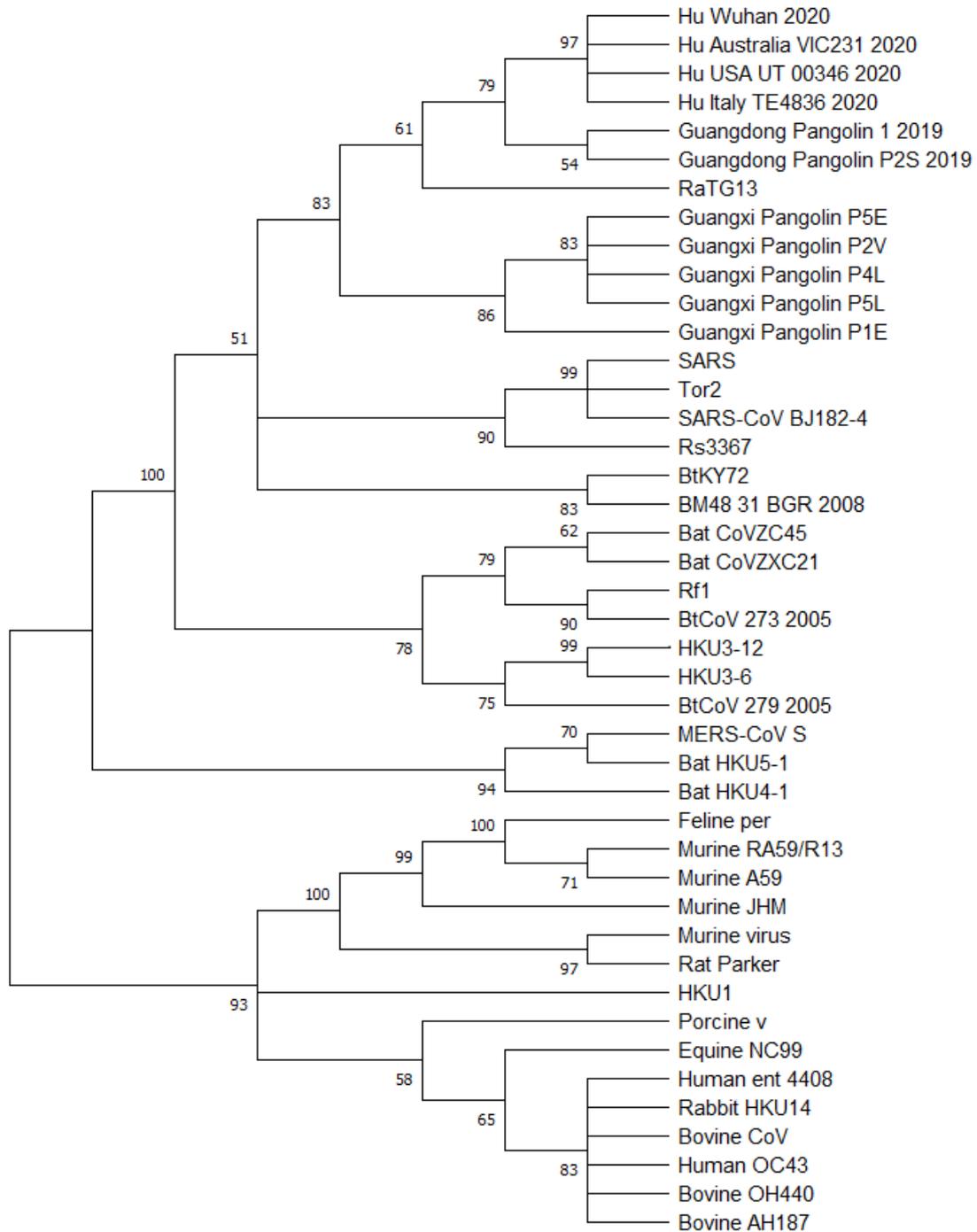

**Supplementary Fig. S16. Phylogenetic tree of the RB domain inferred using RAxML with 100 bootstrap replicates for a group of 43 betacoronaviruses (see S1 Table).**

# COMPLETE LIST OF REFERENCES

Amenta N, Godwin M, Postarnakevich N, John KS. 2007. Approximating geodesic tree distance. *Information Processing Letters*. 103(2):61-65.

Ball GH, Hall DJ. 1965. *ISODATA, a novel method of data analysis and pattern classification*. Menlo Park. Stanford Research Institute.

Ball GH, Hall DJ. 1967. A clustering technique for summarizing multivariate data. *Behavioral Sciences*. 12(2):153-155.

Bansal MS, Burleigh JG, Eulenstein O, Fernández-Baca D. 2010. Robinson-Foulds supertrees. *Algorithms for Molecular Biology*. 5(1):1-12.

Bapteste E, Boucher Y, Leigh J, Doolittle WF. 2004. Phylogenetic reconstruction and lateral gene transfer. *Trends Microbiology*. 12(9):406-411.

Barthélemy JP, McMorris FR. 1986. The median procedure for n-trees. *Journal of Classification*. 3(2):329-334.

Barthélemy JP, Monjardet B. 1981. The median procedure in cluster analysis and social choice theory. *Mathematical Social Sciences*. 1(3):235-267.

Baum BR. 1992. Combining trees as a way of combining data sets for phylogenetic inference, and the desirability of combining gene trees. *Taxon*. 41(1):3-10.

Beiko RG, Harlow TJ, Ragan MA. 2005. Highways of gene sharing in prokaryotes. *Proceedings of the National Academy of Sciences*. 102(40):14332-14337.

Benner P, Bačák M, Bourguignon PY. 2014. Point estimates in phylogenetic reconstructions. *Bioinformatics*. 30(17):i534-i540.

Berry V, Jiang T, Kearney P, Li M, Wareham T. 1999. Quartet cleaning: Improved algorithms and simulations. In: *Algorithms - ESA'99*. Lecture Notes in Computer Science, vol. 1643, Springer, Berlin, Heidelberg.

Berry V, Bininda-Emonds OR, Semple C. 2012. Amalgamating source trees with different taxonomic levels. *Systematic Biology*. 62(2):231-249.

Billera LJ, Holmes SP, Vogtmann K. Geometry of the space of phylogenetic trees. 2001. *Advances in Applied Mathematics*. 27(4):733-767.

Bininda-Emonds OR. 2003. Novel versus unsupported clades: assessing the qualitative support for clades in MRP supertrees. *Systematic Biology*. 52(6):839-848.

Bininda-Emonds OR, editor. 2004. *Phylogenetic supertrees: combining information to reveal the tree of life*. Springer Science & Business Media.

Boc A, Diallo AB, Makarenkov V. 2012. T-REX: a web server for inferring, validating and visualizing phylogenetic trees and networks. *Nucleic Acids Research*. 40(W1):W573-W579.

Boc A, Philippe H, Makarenkov V. 2010. Inferring and validating horizontal gene transfer events using bipartition dissimilarity. *Systematic Biology*. 59(2):195-211.

Bock HH. 2007. Clustering methods: a history of k-means algorithms. *Selected contributions in data analysis and classification*. 161-72.

Boni MF, Lemey P, Jiang X, Lam TT, Perry BW, Castoe TA et al. 2020. Evolutionary origins of the SARS-CoV-2 sarbecoronavirus lineage responsible for the COVID-19 pandemic. *Nature Microbiology*. 5(11):1408-17.

Bonnard C, Berry V, Lartillot N. 2006. Multipolar consensus for phylogenetic trees. *Systematic Biology*. 55(5):837–43.

Bordewich M, Semple C. 2005. On the computational complexity of the rooted subtree prune and regraft distance. *Annals of Combinatorics*. 8(4):409-423.

Bordewich M, Gascuel O, Huber KT, Moulton V. 2009. Consistency of topological moves based on the balanced minimum evolution principle of phylogenetic inference. *IEEE/ACM Transactions on Computational Biology and Bioinformatics*. 6(1):110-117.

Brodal GS, Fagerberg R, Pedersen CN. 2004. Computing the quartet distance between evolutionary trees in time O(nlog n). *Algorithmica*. 38 (2):377–395.

Bryant D, Tsang J, Kearney PE, Li M. 2000. Computing the quartet distance between evolutionary trees. Proc. 11th Annual ACM-SIAM SODA. *Journal of the Society for Industrial and Applied Mathematics*. 9(11):285-286.

Bryant D. 2003. A classification of consensus methods for phylogenetics. *DIMACS series in discrete mathematics and theoretical computer science*. 61:163-84.

Ciccarelli FD, Doerks T, Von Mering C, Creevey CJ, Snel B, Bork P. 2006. Toward automatic reconstruction of a highly resolved tree of life. *Science*. 311(5765):1283-1287.

Caliński T, Harabasz J. 1974. A dendrite method for cluster analysis. Communications Statistics Theory Methods. 3(1):1-27.

Castresana J. 2000. Selection of conserved blocks from multiple alignments for their use in phylogenetic analysis. *Molecular Biology and Evolution*. 17(4):540-552.

Creevey CJ, McInerney JO. 2005. Clann: investigating phylogenetic information through supertree analyses. *Bioinformatics*. 21(3):390-392.

Critchley F, Fichet B. 1994. The partial order by inclusion of the principal classes of dissimilarity on a finite set, and some of their basic properties. In: *Classification and dissimilarity analysis*. Springer, New York, NY, 5-65.

Cotton JA, Wilkinson M. 2007. Majority-rule supertrees. *Systematic Biology*. 56(3):445-452.

Day WH, McMorris FR. 2003. Axiomatic consensus theory in group choice and biomathematics. *Society for Industrial and Applied Mathematics*.

de Queiroz A, Gatesy J. 2007. The supermatrix approach to systematics. *Trends in Ecology and Evolution*. 22(1):34-41.

Dereeper A, Guignon V, Blanc G, Audic S, Buffet S, Chevenet F et al. 2008. Phylogeny.fr: robust phylogenetic analysis for the non-specialist. *Nucleic Acids Research*. 36:W465-W469.

Deza MM, Laurent M. 1997. Geometry of cuts and metrics. *Algorithms and Combinatorics*. Springer-Verlag, Berlin, volume 15.

Dong J, Fernández-Baca D, McMorris FR. 2010. Constructing majority-rule supertrees. *Algorithms for Molecular Biology*. 5(1):2.

Driskell AC, Ané C, Burleigh JG, McMahon MM, O'Meara BC, Sanderson MJ. 2004. Prospects for building the tree of life from large sequence databases. *Science*. 306(5699):1172-1174.

Edgar RC. 2004. MUSCLE: multiple sequence alignment with high accuracy and high throughput. *Nucleic Acids Research*. 32(5):1792-1797.

Felsenstein J. 2013. *Numerical taxonomy*. Springer-Verlag, Berlin Heidelberg, volume 1.

Felsenstein J. 2004. *Inferring phylogenies*. Sunderland (MA): Sinauer Associates, Inc.

Gambette P, Berry V, Paul C. 2012. Quartets and unrooted phylogenetic networks. Journal of bioinformatics and computational biology. 10(04):1250004.

Gambette P, Van Iersel L, Kelk S, Pardi F, Scornavacca C. 2016. Do branch lengths help to locate a tree in a phylogenetic network? *Bulletin of Mathematical Biology*. 78(9):1773-1795.

Gascuel O. 2005. *Mathematics of Evolution and Phylogeny*. Oxford (UK): Oxford University Press, 121-142.

Guénoche A. 2013. Multiple consensus trees: a method to separate divergent genes. *BMC Bioinformatics*. 14(1):46.

Hein J, Jiang T, Wang L, Zhang K. 1996. On the complexity of comparing evolutionary trees. *Discrete Applied Mathematics*. 71(1-3):153-169.

Huson DH, Bryant D. 2005. Application of phylogenetic networks in evolutionary studies. *Molecular Biology and Evolution*. 23(2):254-267.

Jansson J, Shen C, Sung WK. 2013. An optimal algorithm for building the majority-rule consensus tree. In: *Annual International Conference on Research in Computational Molecular Biology*. Springer, Berlin, Heidelberg, 88-99.

Kelly JB. 1972. Hypermetric spaces and metric transforms. *Inequalities II*. Ed. O. Shisha. Academic Press, New York, 149–159.

Kuhner MK, Felsenstein J. 1994. A simulation comparison of phylogeny algorithms under equal and unequal evolutionary rates. *Molecular Biology and Evolution*. 11(3):459-468.

Kumar S, Stecher G, Li M, Knyaz C, Tamura K. 2018. MEGA X: molecular evolutionary genetics analysis across computing platforms. *Molecular Biology and Evolution*. 35(6):1547-1549.

Lam TTY, Jia N, Zhang YW, Shum MHH, Jiang JF, Zhu HC et al. 2020. Identifying SARS-CoV-2 related coronaviruses in Malayan pangolins. *Nature*. 583(7815):282-5.

Li X, Giorgi EE, Marichannegowda MH, Foley B, Xiao C, Kong XP et al. 2020. Emergence of SARS-CoV-2 through recombination and strong purifying selection. *Science Advances*. 6(27):abb9153.

Linz S, Semple C. 2011. A cluster reduction for computing the subtree distance between phylogenies. *Annals of Combinatorics*. 15(3):465.

Lloyd S. 1982. Least squares quantization in PCM. *IEEE transactions on information theory*. 28(2):129-37.

Lord E, Leclercq M, Boc, Diallo AB, Makarenkov V. 2012. Armadillo 1.1: an original workflow platform for designing and conducting phylogenetic analysis and simulations. *PloS One*. 7(1):e29903.

MacQueen J. 1967. Some methods for classification and analysis of multivariate observations. In Proceedings of the fifth *Berkeley symposium on mathematical statistics and probability*. 1(14):281-297).

Maddison DR, Schulz KS, Maddison WP. 2007. The tree of life web project. *Zootaxa*. 1668:19-40.

Maddison DR. 1991. The discovery and importance of multiple islands of most-parsimonious trees. *Systematic Biology*. 40(3):315-328.

Mahajan M, Nimbhorkar P, Varadarajan K. 2009. The planar *k*-means problem is NP-hard. *Lecture Notes Computer Science*. 5431:274-285.

Makarenkov V, Leclerc B. 2000. Comparison of additive trees using circular orders. *Journal of Computational Biology*. 7(5):731-744.

Makarenkov V, Legendre P. 2000. Improving the additive tree representation of a dissimilarity matrix using reticulations. In: *Data analysis, classification, and related methods*. Springer, Berlin, Heidelberg, 35-40.

Makarenkov V, Legendre P. 2001. Optimal variable weighting for ultrametric and additive trees and K-means partitioning: Methods and software. *Journal of Classification*. 18(2):245-271.

Makarenkov V, Mazoure B, Rabusseau G, Legendre P. 2021. Horizontal gene transfer and recombination analysis of SARS-CoV-2 genes helps discover its close relatives and shed light on its origin. *BMC Ecology and Evolution*. 21(1):1-18.

McMorris FR, Meronk DB, Neumann DA. 1983. A view of some consensus methods for trees. In: *Numerical Taxonomy. Proc. NATO Advanced Study Institute on Numerical Taxonomy*. Berlin, Springer Verlag.

McMorris FR, Wilkinson M. 2011. Conservative supertrees. *Systematic Biology*. 60(2):232-238.

Miller E, Owen M, Provan JS. 2015. Polyhedral computational geometry for averaging metric phylogenetic trees. *Advances in Applied Mathematics*. 68:51-91.

Owen M, Provan JS. 2010. A fast algorithm for computing geodesic distances in tree space. *IEEE/ACM Transactions on Computational Biology and Bioinformatics*. 8(1):2-13.

Pan W, Shen X. 2007. Penalized model-based clustering with application to variable selection. *Journal of Machine Learning Research*. 8(41):1145−1164.

Pérez-Losada M, Arenas M, Galan JC, Palero F, Gonzalez-Candelas F. 2015. Recombination in viruses: mechanisms, methods of study, and evolutionary consequences. *Infection, Genetics and Evolution*. 30:296-307.

Prabakaran P, Gan J, Feng Y, Zhu Z, Choudhry V, Xiao X et al. 2006. Structure of severe acute respiratory syndrome coronavirus receptor-binding domain complexed with neutralizing antibody. *Journal of Biological Chemistry*. 281(23):15829-15836.

Ragan MA. 1992. Phylogenetic inference based on matrix representation of trees. *Molecular Phylogenetics and Evolution*. 1(1):53-58.

Robinson DF, Foulds LR. 1981. Comparison of phylogenetic trees. *Mathematical Biosciences*. 53(1-2):131-147.

Rousseeuw PJ. 1987. Silhouettes: a graphical aid to the interpretation and validation of cluster analysis. *Journal of Computational and Applied Mathematics*. 20:53-65.

Sanderson MJ, Purvis A, Henze C. 1998. Phylogenetic supertrees: assembling the trees of life. *Trends Ecology and Evolution.* 13(3):105-109.

Sevillya G, Adato O, Snir S. 2020. Detecting horizontal gene transfer: a probabilistic approach. *BMC Genomics*. 21:106.

Shu Y, McCauley J. 2017. GISAID: Global initiative on sharing all influenza data–from vision to reality. *Eurosurveillance*. 22:30494.

Silva AS, Wilkinson M. 2021. On defining and finding islands of trees and mitigating large island bias. *Systematic Biology*. 70(6):1282–1294.

Snir S, Rao S. 2008. Quartets MaxCut: a divide and conquer quartets algorithm. *IEEE/ACM Transactions on Computational Biology and Bioinformatics*. 7(4),704-718.

St. John K. 2017. The shape of phylogenetic treespace. *Systematic Biology*. 66(1):e83-e94.

Stamatakis A. 2006. RAxML-VI-HPC: maximum likelihood-based phylogenetic analyses with thousands of taxa and mixed models. *Bioinformatics*. 22(21):2688-2690.

Steinley D, Brusco MJ. 2007. Initializing k-means batch clustering: A critical evaluation of several techniques. *Journal of Classification*. 24(1):99-121.

Stockham C, Wang LS, Warnow T. 2002. Statistically based postprocessing of phylogenetic analysis by clustering. *Bioinformatics*. 18(suppl_1):S285-S293.

Sul SJ, Williams TL. 2008. An Experimental Analysis of Robinson-Folds Distance Matrix Algorithms. In: *Esa* 793-804.

Steel M, Rodrigo A. 2008. Maximum likelihood supertrees. *Systematic Biology*. 57(2):243-250.

Sturm KT. 2003. Probability measures on metric spaces of nonpositive curvature. *Communications in Contemporary Mathematics*. 338:357–390.

Swenson MS, Suri R, Linder CR, Warnow T. 2011. SuperFine: fast and accurate supertree estimation. *Systematic Biology*. 61(2):214.

Szöllősi GJ, Daubin V. 2012. Modeling gene family evolution and reconciling phylogenetic discord. *Evolutionary Genomics Methods*. 2:29–51.

Szöllősi GJ, Tannier E, Daubin V, Boussau B. 2014. The inference of gene trees with species trees. *Systematic Biology*. 64(1):e42-e62.

Tahiri N, Willems M, Makarenkov V. 2018. A new fast method for inferring multiple consensus trees using k-medoids. *BMC Evolutionary Biology*. 18(1):48.

Tibshirani R, Walther G, Hastie T. 2001. Estimating the number of clusters in a data set via the gap statistic. *Journal of the Royal Statistical Society B*. 63(2):411-423.

Tseng G. 2007. Penalized and weighted k-means for clustering with scattered objects and prior information in high-throughput biological data. *Bioinformatics*. 23(17):2247–2255.

Wareham HT. 1985. *An efficient algorithm for computing Ml consensus trees*. B.Sc. Honours thesis, Memorial University of Newfoundland, Canada.

Whidden C, Zeh N, Beiko RG. 2014. Supertrees based on the subtree prune-and-regraft distance. *Systematic Biology*. 63(4):566-581.

Wilkinson M, Cotton JA, Lapointe FJ, Pisani D. 2007. Properties of supertree methods in the consensus setting. *Systematic Biology*. 56(2):330-337.

Woodhams MD, Lockhart PJ, Holland BR. 2016. Simulating and summarizing sources of gene tree incongruence. *Genome Biology and Evolution*. 8(5):1299-1315.

Zhang KY, Gao YZ, Du MZ, Liu S, Dong C, Guo, FB. 2019. Vgas: A Viral Genome Annotation System. *Frontiers in Microbiology*. 10:184.